\documentclass[12pt]{article} 
\pdfoutput=1
\input epsf
\usepackage{mathrsfs}
\usepackage{amssymb}
\usepackage{dsfont}
\usepackage[dvips]{graphicx}
\usepackage{color}
\usepackage{amsfonts}
\usepackage{amssymb}
\usepackage{amsmath}
\usepackage{pifont}
\usepackage{bbm}
\usepackage{multirow}
\usepackage{latexsym}
\usepackage{verbatim}
\usepackage{mcite}
\usepackage{cite}
\usepackage{units}
\usepackage[all]{xy}
\usepackage{cancel}
\usepackage{slashed}

\newcommand{\newsection}[1]{
\addtocounter{section}{1} \setcounter{equation}{0}
\setcounter{subsection}{0} \addcontentsline{toc}{section}{\protect
\numberline{\arabic{section}}{{\rm #1}}} \vglue .6cm \pagebreak[3]
\noindent{ \bf  \thesection. #1}\nopagebreak[4]\par\vskip .3cm}
\newcommand{\newsubsection}[1]{
\addtocounter{subsection}{1}\setcounter{subsubsection}{0}
\addcontentsline{toc}{subsection}{\protect
\numberline{\arabic{section}.\arabic{subsection}}{#1}} \vglue .4cm
\pagebreak[3] \noindent{\it \thesubsection.
#1}\nopagebreak[4]\par\vskip .3cm}

\newcommand{\newsubsubsection}[1]{
\addtocounter{subsubsection}{1}
\vglue .4cm \pagebreak[3] \noindent{\it \thesubsubsection.
#1}\nopagebreak[4]\par\vskip .3cm}

\makeatletter
\newcommand{\seclabel}[1]{%
  \@bsphack
  \protected@write\@auxout{}%
     {\string\newlabel{#1}{{\thesection}{\thepage}}}
  \@esphack
  }
\newcommand{\subseclabel}[1]{%
  \@bsphack
  \protected@write\@auxout{}%
     {\string\newlabel{#1}{{\thesubsection}{\thepage}}}
  \@esphack
  }
  \newcommand{\subsubseclabel}[1]{%
  \@bsphack
  \protected@write\@auxout{}%
     {\string\newlabel{#1}{{\thesubsubsection}{\thepage}}}
  \@esphack
  }
\newcommand{\tablabel}[1]{%
  \@bsphack
  \protected@write\@auxout{}%
     {\string\newlabel{#1}{{\arabic{tabnum}}{\thepage}}}
  \@esphack
  }
\makeatother

\renewcommand{\theequation}{\thesection.\arabic{equation}}

\newlength{\extraspace}
\newlength{\extraspaces}
\newcounter{dummy}
\newcommand{\bc}{\begin{center}}
\newcommand{\ec}{\end{center}}
\newcommand{\be}{\begin{equation}
\addtolength{\abovedisplayskip}{\extraspaces}
\addtolength{\belowdisplayskip}{\extraspaces}
\addtolength{\abovedisplayshortskip}{\extraspace}
\addtolength{\belowdisplayshortskip}{\extraspace}}
\newcommand{\ee}{\end{equation}}

\newcommand{\ba}{\begin{eqnarray}
\addtolength{\abovedisplayskip}{\extraspaces}
\addtolength{\belowdisplayskip}{\extraspaces}
\addtolength{\abovedisplayshortskip}{\extraspace}
\addtolength{\belowdisplayshortskip}{\extraspace}}
\newcommand{\ea}{\end{eqnarray}}

\newcommand{\ban}{\begin{eqnarray*}
\addtolength{\abovedisplayskip}{\extraspaces}
\addtolength{\belowdisplayskip}{\extraspaces}
\addtolength{\abovedisplayshortskip}{\extraspace}
\addtolength{\belowdisplayshortskip}{\extraspace}}
\newcommand{\ean}{\end{eqnarray*}}
\newcommand{\baa}{
\addtocounter{equation}{1} \setcounter{dummy}{\value{equation}}
\setcounter{equation}{0}
\renewcommand{\theequation}{\thesection.\arabic{dummy}\alph{equation}}
\begin{eqnarray}
\addtolength{\abovedisplayskip}{\extraspaces}
\addtolength{\belowdisplayskip}{\extraspaces}
\addtolength{\abovedisplayshortskip}{\extraspace}
\addtolength{\belowdisplayshortskip}{\extraspace}}
\newcommand{\eaa}{
\end{eqnarray}
\setcounter{equation}{\value{dummy}}
\renewcommand{\theequation}{\thesection.\arabic{equation}}}

\input epsf

\newcounter{fignum}
\newcounter{tabel}

\newcounter{tabnum}
\begin{document}

%
%

\begin{flushright}
March 2016\\
{\tt UUITP-05/16}\\
\end{flushright}
\vspace{2cm}

\thispagestyle{empty}

\begin{center}
{\Large\bf A new 6d fixed point from holography
 \\[13mm] }

{\sc Fabio Apruzzi$^{1,2,3}$, Giuseppe Dibitetto$^4$ and Luigi Tizzano$^4$}\\[5mm]
{\it $^1$Department of Physics, University of North Carolina, Chapel Hill, NC 27599, USA}\\[5mm]
{\it $^2$CUNY Graduate Center, Initiative for the Theoretical Sciences, New York, NY 10016, USA}\\[5mm]
{\it $^3$Department of Physics, Columbia University, New York, NY 10027, USA}\\[5mm]
{\it $^4$Department of Physics and Astronomy,
     Uppsala university,\\
     Box 516,
     SE-75120 Uppsala,
     Sweden}
\\[15mm]

Abstract:
\end{center}

\noindent We propose a stringy construction giving rise to a class of interacting and non-supersym\-metric CFT's in six dimensions. Such theories may be obtained as an IR
conformal fixed point of an RG flow ending up in a $(1, 0)$ theory in the UV. We provide the due holographic evidence
in the context of massive type IIA on $\textrm {AdS}_{7}\times M_3$, where $M_3$ is topologically an $S^3$.  In particular, in this paper we present a 10d flow solution which may be
interpreted as a non-BPS bound state of NS5, D6 and $\overline{\textrm{D6}}$ branes.
Moreover, by adopting its 7d effective description, we are able to holographically compute the free energy and
the operator spectrum in the novel IR conformal fixed point.

\newpage

\renewcommand{\Large}{\normalsize}

\tableofcontents

\newpage

\newsection{Introduction}

\noindent The discovery of interacting six dimensional superconformal field theories has been a major paradigm shift in our way of thinking about quantum field theories. This new awareness was made possible by a study of various decoupling limits in string theory constructions. Even though 6d SCFT's are local quantum field theories, they resist a Lagrangian description and their low energy degrees of freedoms include tensionless strings. 

Probably the most celebrated example of six-dimensional interacting theory is the supersymmetric $(2,0)$ theory living on the worldvolume of a
stack of coincident M5 branes \cite{Strominger:1995ac}. More precisely, this is a special case of a larger class of $(2,0)$ theories having an ADE classification which can be obtained through geometric engineering in type IIB string theory \cite{Witten:1995zh}. Another prominent example is the 6d $(1,0)$ supersymmetric theory. Recently there has been revived interest in the topic because of the proposed classification of \cite{Heckman:2013pva,Heckman:2015bfa}. Such theories were known for a long time as they also appeared in the context of massive IIA string theory with intersecting NS5--D6--D8 branes \cite{Hanany:1997gh,Brunner:1997gf,Brunner:1997gk}. 

All the SCFT's described by type IIA brane intersections admit a holographic description. The relationship between 6d $(1,0)$ theories and AdS$_7$ solutions of massive type IIA supergravity has been clarified in a series of recent works \cite{Apruzzi:2013yva, Gaiotto:2014lca, Apruzzi:2015zna}. Moreover, the existence of a consistent truncation of massive type IIA supergravity on $\textrm{AdS}_{7}\times S^{3}$ was proven in \cite{Passias:2015gya}. This means that we can use a half-maximal (\emph{i.e.} $16$ supercharges) 7d gauged supergravity restricted to the gravity multiplet as an effective description of the 10d physics. In particular, given an AdS$_7$ solution of gauged supergravity, its string theory lift can be constructed explicitly. 

All the AdS$_7$ solutions of half-maximal gauged supergravity have been classified in \cite{Dibitetto:2015bia}. A striking outcome of this analysis is the presence of non-supersymmetric solutions. More surprisingly, a computation of the full perturbative spectrum of one of the non-supersymmetric vacua confirms an absence of tachyons. In light of these results, the aim of this paper is to give a complete \emph{holographic interpretation of this non-supersymmetric $\textrm{AdS}_7$ solution}. In particular we will provide \emph{evidence for the existence of a new class of non-supersymmetric CFT$_6$ dual fixed points.}

The existence of 6d superconformal fixed points in string theory is a well established fact. To the best of our knowledge a string theoretical description of a fully non-supersymmetric 6d fixed point is still unknown. The goal of this paper is to give evidence for the existence of a class of non-supersymmetric CFT's, and we also pursue a first step in the string theoretic understanding of these 6d CFT's.

In this work we address the problem in terms of holographic RG flows \cite{Freedman:1999gp}. We use a static non-BPS domain wall (DW) solution, discovered some time ago in \cite{Campos:2000yu}, interpolating between the supersymmetric and non-supersymmetric AdS$_7$ vacua. This is interpreted as the first evidence of the existence of a 6d RG flow connecting the supersymmetric solution in the UV with the non-supersymmetric one in the IR. Additionally, we construct the corresponding 10d flow which reconciles the 7d effective description with the string theoretical susy breaking process. 

The second evidence comes from a study of holographic quantities associated with the new CFT$_6$. We are able to check the free energy of the corresponding AdS$_7$ solution and its large $N$ behavior. Following \cite{Girardello:1998pd}, we are able to \emph{test} positively the validity of the 6d ``a-theorem"\footnote{Despite the holographic evidence \cite{Myers:2010tj}, there is still not a general proof of the a-theorem for six dimensional CFT's. In this case the dilaton effective action analysis turns out to be inconclusive \cite{Elvang:2012st,Maxfield:2012aw}. See also \cite{Grinstein:2014xba} for further complications. Finally, the classification of supersymmetric deformations of 6d SCFTs has been carried out in \cite{Cordova:2016xhm, Louis:2015mka}. } in our 6d RG flow using the identification between central charge and stress energy tensor of the 7d supergravity DW solution. Finally we are able to track the evolution of conformal dimensions for the operator spectrum and to identify which operator is responsible for triggering the flow.

The third and final evidence comes from the susy breaking description at the level of brane constructions in string theory. We propose a mechanism, along the line of \cite{Kachru:2002gs, Argurio:2006ny, Argurio:2007qk} which involves an explicit use of $\overline{\textrm{D6}}$ branes in the original massive IIA brane system together with a jump in the fluxes in order to keep the total D6 brane charge unaffected. This implies that the charges of supergravity at infinity are not modified and we obtain a faithful description of holographic states in the same theory. The presence of $\overline{\textrm{D6}}$'s is an explicit source of susy breaking which generates a non-supersymmetric metastable state in the tensor branch of the original $(1,0)$ theory. This metastability ends at the new interacting fixed point, where the distance between NS5 branes goes to zero and conformal symmetry is restored. This precisely corresponds to the process described within the holographic RG flow. 

Working with AdS/CFT outside the supersymmetric realm is certainly more subtle. A common problem in these cases is that non-supersymmetric solutions usually suffer from gravitational instabilities. An interesting aspect of the class of non-supersymmetric $\textrm{AdS}_7$ solutions of this paper is their stability both at a perturbative and at a non-perturbative level. We provide a confirmation for our claim by reformulating a positive energy theorem type of argument. The fact that the gravity solution we are using is free from these instabilities makes it a sensible tool to test its dual non-supersymmetric $\textrm{CFT}_6$.   

A final comment concerns the role of $\overline{\textrm{D6}}$ branes in our construction. Here we provide a qualitative check of the tension of gravity solutions at both end points of the DW. This check confirms our expectation about the possibility of tracing the presence of susy breaking anti-branes directly in the gravity solution. We also make some preliminary comments about the full near-horizon solution of non-BPS NS5--D6--D8 brane intersections in massive IIA, which is an unknown terrain. Of course, with this analysis we only scratched the surface. Given the new and intriguing potential applications to holography, it is a major challenge to refine our understanding of brane solutions in massive IIA string theory.  

This paper is organized as follows: In section~\ref{sec:6dtheory}, we first review the stringy construction of 
$(1,0)$ SCFT's and then discuss susy breaking and tachyon condensation as a mechanism for producing novel non-supersymmetric 
conformal fixed points in 6d.
Subsequently, in section~\ref{sec:10Dgravity} we move to the description of those non-supersymmetric $\textrm{AdS}_7$
vacua in massive type IIA supergravity that provide gravity duals of the aforementioned  $\textrm{CFT}_6$. Moreover we give a 10d flow solution interpolating between the supersymmetric and the non-supersymmetric vacua. Finally, we show how all the above information is captured by a 7d gauged supergravity.
In section~\ref{sec:CFT6}, adopting this 7d effective desciption, we perform the holographic computation of the free energy and determine the operator spectrum of its dual $\textrm{CFT}_6$. 
In section~\ref{sec:stability}, we first comment on the stability of the non-supersymmetric AdS vacuum, both at a perturbative and non-perturbative level. 
Later we include a computation of the effective tension of the sourcing brane system allowing us to argue for the presence of anti-branes. 
Finally in section~\ref{sec:conclusion}, we conclude by adding some physical remarks and some hints at future directions to be pursued.
Some extra technical material concerning supergravity may be found in appendices~\ref{App:MIIA} and \ref{App:Gauged_Sugra}.

\newpage

\newsection{6d $(1,0)$ SCFT's and susy breaking}\seclabel{sec:6dtheory}
\noindent As predicted by string theory, there are many different manifestations of six-dimensional superconformal field theories. In this section we will focus on 6d $(1,0)$ SCFT's obtained through NS5--D6--D8 brane configurations in massive IIA theory. The reason for this choice is that this class of theories is particularly useful for our holographic purposes, although this fact will be clearer only in the next section. Moreover, the brane construction is very efficient to visualize our proposal for a susy breaking process which relies on the explicit presence of anti-branes. 
\newsubsection{Brane engineering}
\subseclabel{sec:braneeng}
\noindent We begin by reviewing an important class of supersymmetric 6d theories which have known gravity duals. These were originally introduced by Hanany and Zaffaroni in \cite{Hanany:1997gh}. See also \cite{Brunner:1997gf,Brunner:1997gk}. 

The basic idea is to start from the Hanany-Witten (HW) \cite{Hanany:1996ie} setup, whose basic ingredients are NS5, D5 and D3 branes.  We place a D3 brane with worldvolume $(0126)$ together with an NS5 brane with worldvolume filling $(012345)$ and a D5 brane on $(012789)$. Performing three different T-dualities respectively in the directions $(345)$ leads to the following brane configuration:  
\begin{center}
\begin{tabular}{c|c|c|c|c|c|c|c|c|c|c}
         & 0 & 1 &  2 & 3 & 4 & 5 & 6 & 7 & 8 & 9 \\
       \hline  \hline 
NS5 & $\times$ &  $\times$ &  $\times$ &  $\times$ &  $\times$ & $\times$ &    &       &   &   \\
\hline
D8    & $\times$ &  $\times$ &  $\times$  &  $\times$ &  $\times$ &  $\times$ &  &  $\times$  &   $\times$  &  $\times$   \\
\hline
D6    & $\times$ &  $\times$ & $\times$ &  $\times$ &   $\times$ &   $\times$  &   $\times$ &   &      &    \\
\end{tabular}
\end{center}
\begin{figure}[h]
\begin{center}
\includegraphics[scale=0.38]{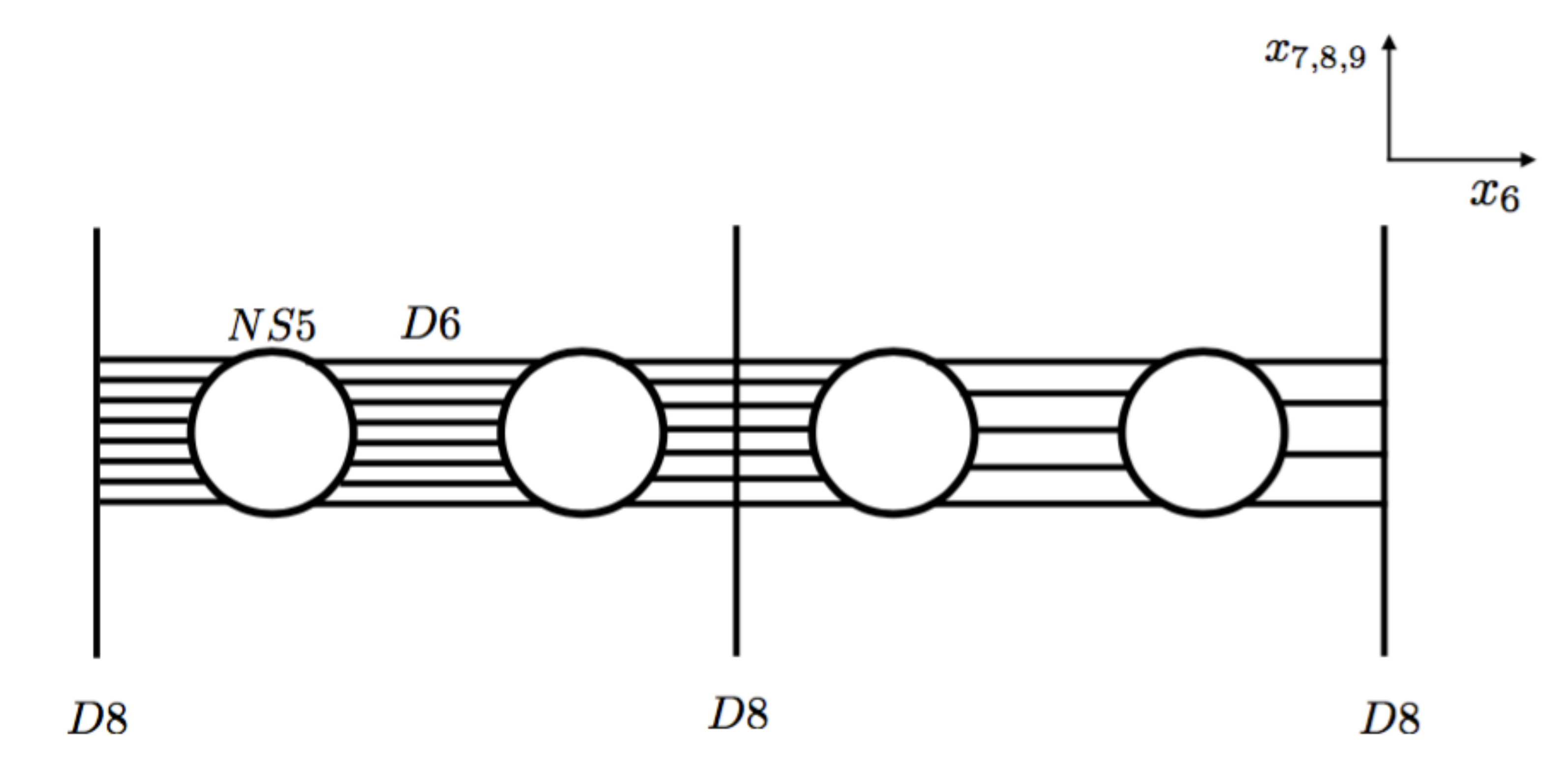}
\caption{\it In this picture, the vertical lines are D8 branes extending along the $x^{7,8,9}$ dimensions, they look like walls separating the six space-time dimensions in common between the NS5 and D6 branes. The fat dots are the stacks of NS5 branes and the horizontal lines correspond to D6 branes extending along $x^6$.}
\label{fig:system1}
\end{center}
\end{figure}
This system can be visualized in figure~\ref{fig:system1}. The D6 branes stretch along the $x_6$ direction, the D8 branes look like vertical walls and the NS5's are localized at points along the $x_6$ line on which the D6's are assumed to end. The physical intuition about multiple branes configurations is that the low energy dynamics is characterized by the lowest dimensional brane in the setup. In this case, we have a 6d supersymmetric theory with a $(1,0)$ supersymmetry.  As we can see from the brane system, there is an $\textrm{SO}(3)$ symmetry on the $789$ directions which represents the $\textrm{SU}(2)$ R-symmetry of the $(1,0)$ theory. A crucial difference with respect to the HW setup is that here the NS5 branes give rise to a tensor multiplet that contributes to the spectrum of the 6d theory.
The worldvolume theory of the D8 branes has some infinite directions, which may be seen as a frozen background by the 6d theory thus generating a global symmetry.

In figure~\ref{fig:system1}, we have $N$ NS5 branes at various position $x_i^6$, $k_i$ D6 branes stretched within the intervals between to consecutive locations of NS5's. We denote by $w_i$  the number of D8 branes in the $i-(i+1)$ interval. This system has a gauge symmetry
\be
\prod_{i}^{N-1} \textrm{U}(k_i)
\ee
inherited by the reduction of the D6 branes' worldvolume theory along $x^6$. As in the HW setup the distances between NS5 branes are associated to the gauge couplings of each factor. Namely, for the $i$-th factor we have the following relation
\be
\frac{1}{g_i^2} \ = \ \phi^{6}_{i+1} \, - \, \phi^{6}_{i}
\ee
where $\phi_i^6$ denotes the scalar field in the $(1,0)$ tensor multiplet associated with the position $x_i^6$.  Notice that the presence of D8 branes in the theory implies that we are working in massive IIA with $F_{(0)} \,=\,\pm k$ for very large positive and negative $x^6$. Therefore, the M-theory lift of this configuration is still not known.  \\
 
All the theories described here have an analogous of the Coulomb branch, in 6d called tensor branch, parametrized by real scalars in the tensor multiplet, which can be interpreted as the relative positions of the NS5's along $x^6$. If we go to the origin of the tensor branch, where the relative distances between the solitonic branes is zero, we expect the appearance of tensionless strings as new degrees of freedom in the theory. Gauge factors associated with pairs of coinciding NS5 branes are strongly coupled in this limit. It is widely believed that it is not possible to treat this kind of interactions with the usual paradigm of a continuum Lagrangian description, \emph{i.e.} as perturbations a Gaussian fixed point. Such a phenomenon is instead interpreted as a sign of RG interacting fixed points.       
For this reason, these theories are usually referred to  as ``non-Lagrangian''. Thanks to the increasing interest in such theories, an incredible amount of novel examples thereof has been found in the recent years. 

Many new properties of the six-dimensional $(1,0)$ theories we described so far were recently pointed out in the work of \cite{Gaiotto:2014lca}.
The idea there was to analyze in detail the boundary conditions for the D6 branes ending on the stack of D8's in analogy with the three dimensional case worked out in \cite{Gaiotto:2008sa}. It turns out that separating D8 branes create new Higgs branch deformations for the 7d gauged theory living on the D6 brane segments stretched between D8 branes. In this way, starting from the interacting theory and performing Higgs and tensor branch deformations, the Hanany-Zaffaroni brane system is obtained as a result.

When looking at the theory in the tensor branch, there are certain general properties of the underlying brane system that should always be taken into account. Moving a D8 brane along the $x^6$ direction and passing trough an NS5 brane will create or annihilate the appropriate number of D6 branes as dictated by the HW effect. Moreover, the anomaly cancellation condition for six-dimensional theories requires that $N_f = 2N_c$ at each node in the quiver. In the brane system, this requirement can be expressed as
\be\label{eq:charge}
n_{\rm D6,left} \, - \, n_{\rm D6,right} \ = \ n_0 \ = \ 2\pi F_{(0)} \ .
\ee
This follows from the Bianchi identity (BI) $dF_{(2)}\,-\,H_{(3)}\,F_{(0)} \, = \, \delta_{\rm D6}$, integrated around the D6's before and after an NS5.

\newsubsection{Susy breaking}

\noindent At present, all known examples of interacting CFT's in six dimensions are supersymmetric and arise
from decoupling limits of string constructions. In this section we conjecture a possible consistent picture for susy breaking of the six-dimensional $(1,0)$ theories studied in this paper. We propose a use of susy-breaking anti-branes in the spirit of \cite{Kachru:2002gs, Argurio:2007qk} to argue in favor of the existence of a new class of non-supersymmetric 6d CFT's. Our considerations here are further supported by a detailed study of the dual gravitational solutions that will be presented in the next section.   

The starting point is the brane diagram of the six-dimensional $(1,0)$ theory in the tensor branch as presented in figure~\ref{fig:system1}. In this setup we assume that the D8 branes are far away and do not play an active role. The NS5's are separated and look like lines on the $\left(x^5,x^6\right)$ plane spanned by the D6 branes. 
\begin{figure}[h]
\begin{center}
\includegraphics[scale=0.38]{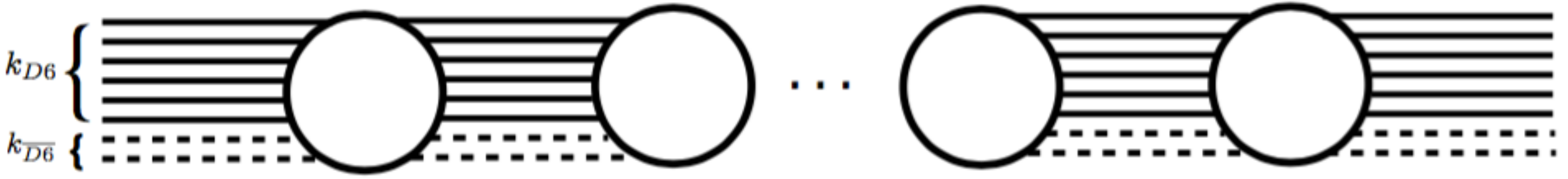}
\caption{\it The shaded horizontal segments symbolize $\overline{\textrm{D6}}$ branes extending between NS5 branes.}
\label{fig:system2}
\end{center}
\end{figure}

At this pont we would like to review a more familiar situation. In four dimensions in fact there is a very similar brane system which consists of NS5--D4--D6 branes. There, it is possible to move two NS5 branes across each other. In the process, some $\overline{\textrm{D4}}$ branes are generated in the interval between the two NS5 branes and they annihilate against D4 branes sitting on top of them. This is the brane picture of \emph{Seiberg duality}, which was originally proposed in \cite{Elitzur:1997fh} and reinterpreted in \cite{Argurio:2007qk} using anti-branes.

The idea behind \cite{Argurio:2007qk} is that of generating a non-zero meson vev in the \emph{electric theory} by moving the D4 branes away from their position. In this way, if we perform a Seiberg duality, the brane system undergoes a net creation of $\overline{\textrm{D4}}$ branes which do not annihilate because the D4 branes are displaced from their original position.  This means that the \emph{magnetic theory} is non-supersymmetric and its brane system is the string theory representation of the ISS vacuum \cite{Intriligator:2006dd}. A final important remark is that in type IIA there is an alternative way of obtaining the same final situation; namely by adding D4/$\overline{\textrm{D4}}$ pairs to all the intervals (where the D4 branes are stretched between the NS5's) except the one between the two interchanged NS5's.

In six dimensions things are more complicated and in fact we do not know if there exists a dynamical process along the line of \cite{Argurio:2007qk}, which could explain the susy breaking. Even though we do not have such a dynamical description, we propose the following brane engineering to obtain a final non-supersymmetric theory. Starting from the brane system in figure~\ref{fig:system1}, we add $\overline{\textrm{D6}}$ branes to it while having fluxes jump in the gravity description, such in a way that the full D6 brane charge \eqref{eq:charge} remains unaffected. This mechanism is depicted in figure~\ref{fig:system2}. The simple argument behind this statement is that, at the level of the brane charges, it is simply impossible to distinguish between the original system and the one in presence of the $\overline{\textrm{D6}}$ branes. This can be seen in the following way. First of all, excluding the presence of D8 branes, it is known that gravity solutions supported by brane sources must satisy 
\be\label{eq:branecharge2}
k_{D6} \, - \, k_{\overline{D6}} \ = \ N \,F_{(0)}\ ,
\ee
where $k_{D6}$ and $k_{\overline{D6}}$ are the number of D6 and $\overline{\textrm{D6}}$ branes respectively and $N$ is the number fo NS5's. The \eqref{eq:branecharge2} may be regarded as a consequence of the BI for $F_{(2)}$. At this point, it is clear that situations with different numbers of D6 and $\overline{\textrm{D6}}$ branes can be described in this way, provided that a jump in the fluxes on the rhs of \eqref{eq:branecharge2} is accordingly taken into account.

The use of anti-branes to construct non-supersymmetric vacua was pioneered in \cite{Kachru:2002gs} and applied in many other contexts by many authors. The hallmark of this approach is that in order for the new configuration to describe states in the same theory, the supergravity charges at infinity should be unchanged. Notice that this requirement nicely fits with our proposal for the susy breaking within the 6d $(1,0)$ theory.

In the conventional language of susy-breaking literature, we just obtained a meta-stable non-supersymmetric version of the 6d $(1,0)$ theory in the tensor branch. The meta-stability here is triggered by the explicit presence of anti-branes in the brane construction of the gauge theory. It would be nice to have a field-theoretic description of this process even though the absence of a Lagrangian formulation seems to be a great obstacle in this case.  

Even though the use of anti-branes might generically leads to gravitational instabilities, here one should not be too worried about this stability issue for two reasons. Firstly, the theory that we are now considering has not quite reached yet a CFT point. Secondly, the gauge-theoretic meta-stability does not reflect an actual instability of the gravity solution that we are going to describe in the next section. The simultaneous presence of D6 and $\overline{\textrm{D6}}$ branes creates a tension in the system that triggers an annihilation process which is eventually resolved when the distance between the solitonic NS5 branes tends to zero. Indeed, we know that this limit is very special because it corresponds to the emergence of a new interacting conformal fixed point. 

Let us summarize our main idea. A non-supersymmetric meta-stable state of a 6d $(1,0)$ theory can be resolved in the limit where the distance between the NS5 branes tends to zero. This is the sign of a new non-supersymmetric 6d conformal fixed point, which can be reached in the IR through an RG flow. We claim that such a fixed point exists and that it can be studied in the context of the $\textrm{AdS}_7/\textrm{CFT}_6$ correspondence.

\newsubsection{Annihilation of D6's and $\overline{\textrm{D6}}$'s between NS5 branes}
\subseclabel{sec:tachy}
\noindent We now provide more details about the flow between the two conformal fixed points. As already anticipated, the process that should lead to a non-supersymmetric fixed point in the IR involves a move into the tensor branch, where the NS5 branes are separated along the $x^6$ direction. While separating the NS5's, the right amount of D6's and $\overline{\textrm{D6}}$'s is created, such that the charge conservation constraint \eqref{eq:branecharge2} is fulfilled. The presence of $\overline{\textrm{D6}}$'s between two NS5 nodes is quite crucial here. Notice that, since the supersymmetric fixed point in the tensor branch moduli space is unique, whenever reaching a new fixed point in the IR, it will necessarily be non-supersymmetric. 

Let us give a brief description of the RG flow that reaches the new fixed point in the IR in terms of brane configurations. This non-supersymmetric flow involves a presence of branes and anti-branes stretching between the NS5's. For simplicity we focus on two NS5's separated by coincident $k_{\rm D6}$ and $k_{\overline{\rm D6}}$ stretching along the interval in between in the $x^6$ direction. Some of the 6-branes tend to annihilate pairwise because of the presence of open-string tachyons \cite{Sen:1998sm,Sen:1998rg,Sen:1999mg}. Eventually, these tachyons condense and the brane system decays to a non-supersymmetric fixed point. We claim that this non-trivial effect is due to the presence of the NS5's. To illustrate our idea we focus on a D6/$\overline{{\rm D6}}$ pair in an NS5 background. As shown in \cite{Kutasov:2004dj, Kutasov:2004ct, Kluson:2004xc}, this system can be described with a tachyon effective action of the form  
\begin{equation}
S_{\tau}\,=\,-T_6\,\int d^{7}x  \frac{\mathcal V(\tau)}{\sqrt{h(R)}} \sqrt{{\rm det}(\eta_{\mu \nu} + h(R)\, \partial_{\mu}R\,\partial_{\nu}R + \partial_{\mu}\tau\, \partial_{\nu}\tau)\,} \ , 
\end{equation}
where $\tau$ is a ``tachyon" field,  $\eta_{\mu \nu}$ is the metric on the NS5 worldvolume theory and $R$ is a field coinciding with the distance between two NS5's. The dynamics of the center of mass is decoupled, and $h$ is then given by the harmonic function of the NS5 background
\begin{equation}
h(R) \ = \ 1\,+\,\frac{1}{R^2} \ .
\end{equation}
The total effective potential is given by
\begin{equation} \label{eq:NS5pot}
\mathcal V(\phi, \tau)\,=\,\frac{T_6}{\sqrt{h(R)}\, {\rm cosh}\left(\frac{\tau}{\sqrt{2}}\right)\,}\ .
\end{equation}
\begin{figure}[h]
\begin{center}
\includegraphics[scale=0.5]{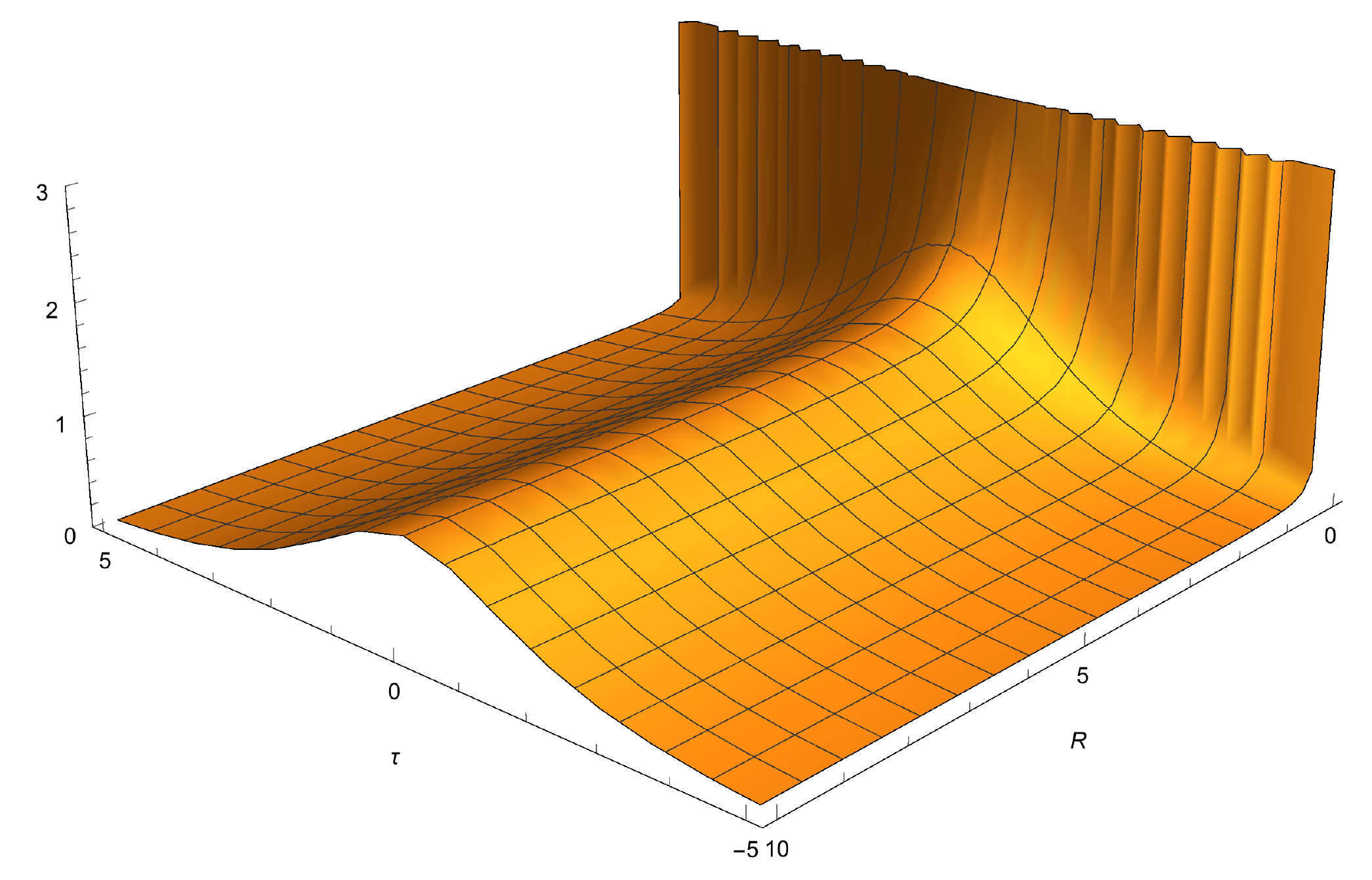}
\caption{\it Plot of potential \eqref{eq:NS5pot}.}
\label{fig:NS5pot}
\end{center}
\end{figure}

\begin{figure}[h!]
\begin{center}
\includegraphics[scale=0.6]{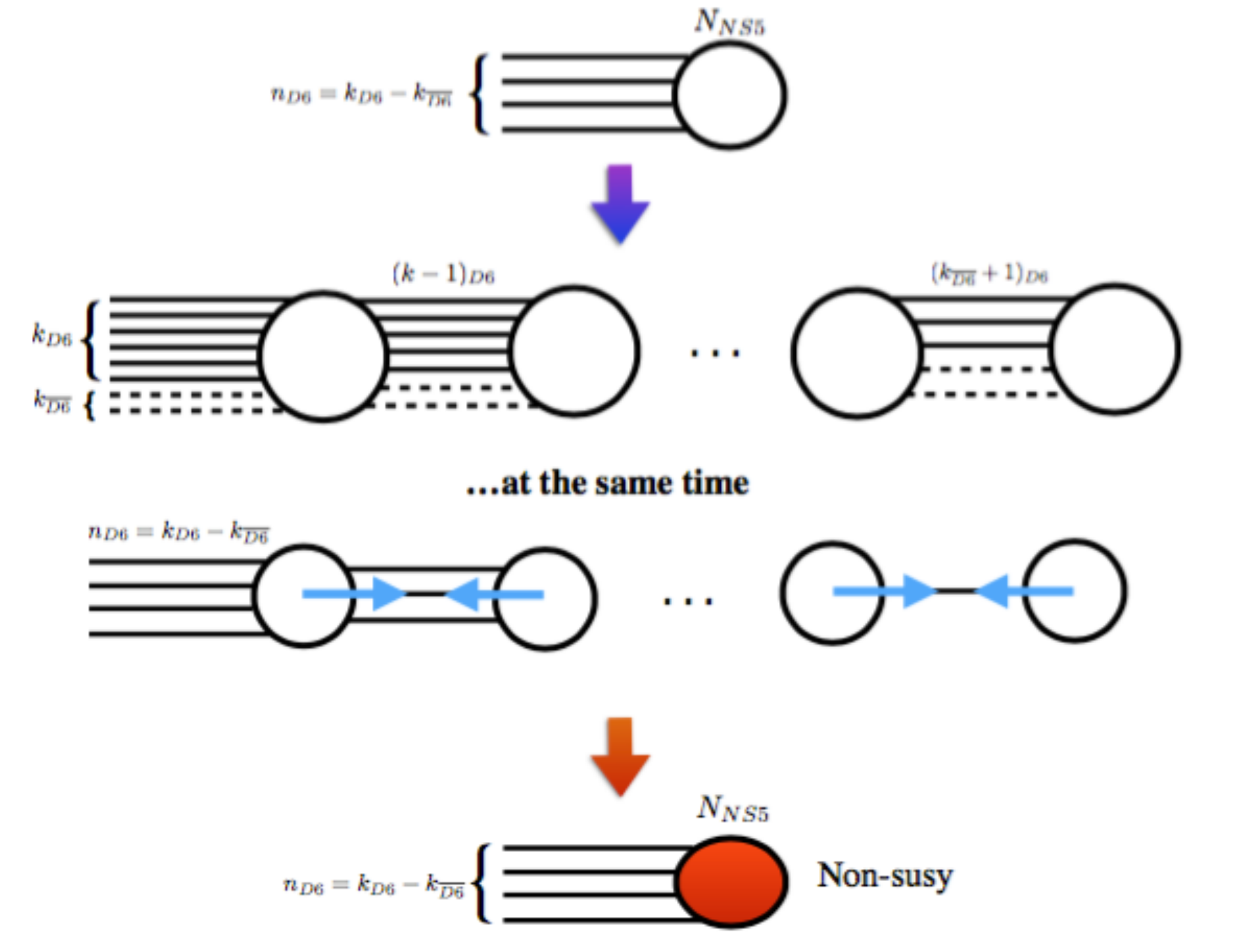}
\caption{\it An artist's impression of the non-susy RG flow for a specific example of $(1,0)$ theory. The vertical arrows indicate the evolution of the flow, whereas the horizontal blue arrows describe the attractive force between NS5 branes driven by tachyon condensation. Annihilation and attraction of the NS5 happen to be a simultaneous process as indicated by the potential \eqref{eq:NS5pot}.}
\label{fig:system3}
\end{center}
\end{figure}

The potential $\mathcal{V}$ provides a measure of the interaction between the distance of the NS5 and the tachyon field. In an effective field theory description this would be identified with a coupling of the theory. When $\tau\rightarrow \pm \infty$ and $R\rightarrow 0$, $\mathcal{V}$ goes to zero and the coupling disappears. It is quite remarkable that the limit where the tachyon coupling is weak coincides with the emergence of the conformal fixed point. We interpret this phenomenon as a clear indication that the presence of $\overline{\textrm{D6}}$ is deeply related to the existence of a new non-superysmmetric IR fixed point. The mechanism through which the $\overline{\textrm{D6}}$'s trigger the RG flow is sketched in figure~\ref{fig:system3}.  Moreovoer, the potential \eqref{eq:NS5pot} and its behavior in figure \ref{fig:NS5pot} suggest that the process of annihilation and the appearance of the attractive forces between the NS5 branes happen simultaneously. 

It would be very interesting to repeat this analysis in a more complicated background. In particular the brane/anti-brane pair should be studied in a background that incorporates the effect of original supersymmetric D6 segments separating the NS5's.
\newpage

\newsection{$\textrm{AdS}_7$ vacua and interpolating solutions}\seclabel{sec:10Dgravity}

\noindent In the previous section we have proposed an intuitive stringy construction giving rise to a dynamical susy breaking mechanism within an NS5--D6--D8 system. This allowed us to argue in favor of  the existence of non-supersymmetric conformal fixed points in 6d which are connected to $(1,0)$ theories via RG flows.  

The aim of this section will be that of supporting the above intuition by constructing its holographic evidence in the context of massive type IIA supergravity. This theory is known to admit supersymmetric
$\textrm{AdS}_7$ vacua thanks to the work of \cite{Apruzzi:2013yva} and \cite{Apruzzi:2015zna}, where the general supersymmetric IIA system of equations was analyzed using ``pure spinor'' techniques.

Subsequently, in \cite{Passias:2015gya} it was shown that there exists a consistent truncation of massive type IIA supergravity on $\textrm{AdS}_{7}\times S^{3}$ yielding a half-maximal 7d gauged supergravity restricted to the gravity multiplet as an effective description. This allows us to explicitly construct the 10d lift of all $\textrm{AdS}_{7}$ solutions found in \cite{Dibitetto:2015bia} in
the gauged supergravity context. These will include, in particular, a non-supersymmetric one\footnote{A hint at the possible existence of interesting non-supersymmetric AdS$_7$ solutions of massive IIA was first given in \cite{Junghans:2014wda}, though their relation to the ones analyzed here remains to be studied in detail and clarified.}. Moreover, exploiting the 7d effective description, we will be able to compute the full perturbative spectrum of
closed string excitations, which turns out to be tachyon-free even in the non-supersymmetric solutions.

Finally, we present a static non-BPS DW solution interpolating between the supersymmetric and non-supersymmetric AdS vacua providing us with the due evidence of the existence of a 6d RG flow connecting the supersymmetric fixed point in the UV with the non-supersymmetric one in the IR. We will also construct the corresponding 10d flow. This will turn out to describe a non-BPS, but \emph{extremal} NS5--D6--$\overline{\textrm{D}6}$ system admitting an $\textrm{AdS}_{7}\times S^{3}$
near-horizon geometry.

\newsubsection{The 10d flow Ansatz} \subseclabel{sec:DWans}
\noindent In order to study the lift of a non-BPS DW solution, we will cast an explicit 10d \emph{Ansatz} for the metric and the forms. In general this \emph{Ansatz} will be of cohomogeneity 2 (\emph{i.e.} all the fields depend on two coordinates $(z,r)$). However, we will use a certain specialization of the \emph{Ansatz} given in \cite{Passias:2015gya}, which suitably splits the $z$ \& the $r$ dependence.
Despite that such a splitting will not consist in a factorized $(z,r)$-dependence in the original \emph{Ansatz}, it will succeed in converting the original field equations into ODE's.

The \emph{Ansatz} for the 10d metric reads,
\begin{equation} \label{eq:DWmetric}
\hspace{-1mm}
ds^2_{10} \,=\, \frac{\kappa ^2}{8}  X(z)^{-\frac{1}{2}} e^{2 A(r)}\left(dz^2\,+\,e^{2a(z)}ds^2_{6}\right)+ X(z)^{\frac{5}{2}}\left(dr^2+\frac{e^{2 A(r)} \left(1-x(r)^2\right)}{16 \, w(z,r)} d\Omega^2_{(2)}\right),
\end{equation}
where $\kappa$ is a constant, $ds^2_{6}$ \& $d\Omega^2_{(2)}$ denote the flat $\textrm{Mkw}_{6}$ and the round unit $S^{2}$ metrics, respectively, $w(z,r)\,\equiv\, x(r)^2\,+\,X(z)^5 \left(1-x (r)^2\right)$
and the functions $a$ \& $X$ only depend on the DW coordinate $z$. 
The 10d dilaton is given by
\begin{equation}
\Phi\,=\,\Phi_{0}(r)\,+\,\frac{1}{2} \log \left(\frac{X(z)^{\frac{5}{2}}}{w(z,r)}\right) \ .
\end{equation}
We will see that $\Phi_{0}(r)$ will correspond to the dilaton profile of the supersymmetric solution found in \cite{Apruzzi:2013yva} and \cite{Apruzzi:2015zna}. The NS-NS 2-form potential is given by 
\begin{equation}
B_{(2)}\,=\,\frac{e^{2 A(r)} \, x(r) \, \sqrt{1-x (r)^2}}{16 \, w(z,r)} \, {\rm vol}_{S^2} \ ,
\end{equation}
and the $H_{(3)}\,=\,dB_{(2)}$. Finally, the \emph{Ansatz} for the $F_{(2)}$ flux is
\begin{equation}
F_{(2)}\,=\,\sqrt{1-x(r)^2}\,\left( \frac{x(r) \, e^{2 A(r)}}{\sqrt{2}  \kappa^3 \, w(z,r)}\, F_{(0)}  - \frac{1}{4} e^{A(r)-\Phi_{0}(r) }\right) {\rm vol}_{S^2} \ .
\end{equation}
In what follows we will analyze the constraints imposed by the set of field equations and BI in appendix~\ref{App:MIIA} on our 10d \emph{Ansatz} of the form of \eqref{eq:DWmetric}. For such a case, we will see
how the original set of PDE's is turned into a set of ODE's constraining the functions $\Phi_{0}(r)$, $A(r)$, $x(r)$ as well as $a(z)$ \& $X(z)$.

\newsubsection{Asymptotic AdS vacuum Solutions} \subseclabel{sec:asympsol}
\noindent When looking for AdS vacuum solutions, one should further specialize the \emph{Ans\"atze} in section~\ref{sec:DWans} to constant profiles for the scalar $X$, in order to recover the 
$\textrm{AdS}_7$ behavior at $z\rightarrow \pm \infty$. In particular, one sets 
\begin{equation}\label{const_X}
\begin{array}{lccclc}
X(z) \ = \ X & , & & & a(z)\ = \ \dfrac{z}{\ell}\ , \qquad \textrm{where} \qquad \ell\,=\, \dfrac{2 \sqrt{30} \, X^4}{\kappa \, \sqrt{8 X^{10}+8 X^5-1}} & .
\end{array} 
\end{equation}
Plugging this into the 10d equations \eqref{eq:dileom}, \eqref{10s_Einstein}, \eqref{eq:fluxeom} and \eqref{eq:bianchi} in appendix~\ref{App:MIIA}, we get that the full set of 10d field equations and BI is
equivalent to the following set of ODE's  
\be\label{eq:oder}
\left\{
\begin{array}{lclc} 
\Phi_{0}'(r) & = & \dfrac{1}{4} \dfrac{e^{-A(r)}}{\sqrt{1-x(r)^2}} \, \left(12 \, x(r) \, + \, \left(2x(r)^2-5\right)\,F_{(0)} \, e^{A(r)+\Phi_{0}(r)}\right) & , \\[4mm]
x'(r) & = & -\dfrac{1}{2} e^{-A(r)}\sqrt{1-x(r)^2} \, \left(4 \, + \, x(r) \, F_{(0)}\, e^{A(r)+\Phi_{0}(r)}\right) & ,\\[3mm]
A'(r) & = & \dfrac{1}{4} \dfrac{e^{-A(r)}}{\sqrt{1-x(r)^2}} \, \left(4\,x(r) \, - \, F_{(0)} \, e^{A(r)+\Phi_{0}(r)}\right) & ,
\end{array}\right.
\ee
together with the following algebraic condition on $X$:
\begin{equation} \label{eq:algcX}
2 X^{10} \,- \, 3 X^5 \, + \, 1 \ = \ 0 \ .
\end{equation}
The ODE's \eqref{eq:oder} are exactly the same ODE's found in \cite{Apruzzi:2013yva}. Moreover, \eqref{eq:algcX} is solved for the following two values of $X^5$
\begin{equation}
\begin{array}{lccclc}
X^5 \ = \ 1 & , & & &  X^5 \ = \ \dfrac{1}{2} & ,
\end{array}
\end{equation}
where the first value corresponds to the supersymmetric solution at $z\rightarrow +\infty$ and the second to the non-supersymmetric one at $z \rightarrow -\infty$.

\newsubsubsection{Supersymmetric $\textrm{AdS}_7$ Solutions} \subsubseclabel{sec:susyads}
\noindent The supersymmetric solution that one asymptotes to as $z\rightarrow +\infty$ is of the type of \cite{Apruzzi:2013yva}. All the AdS$_7$ solutions have a ten-dimensional metric which can be written as a warped product
\begin{equation}
ds^2_{10}\,=\, e^{2A(r)} ds^2_{{\rm AdS}_7} \,+\, ds^2_{M_3}\ .
\end{equation}
Moreover, since $(1,0)$ theories have an SU(2) R-symmetry, all supersymmetric solutions of the above type need to have an $M_3$ constructed from a fibration of a round $S^2$ over a one dimensional space:
\begin{equation}\label{eq:met-r}
ds^2_{M_3} \, = \, dr^2 \, + \, \frac1{16}e^{2A(r)}\,(1-x(r)^2)\,d\Omega_{(2)}^2 \ , 
\end{equation}
where the function $x$ is related to the volume of the $S^2$ and the dilaton $\Phi_{0}$ is a function of $r$ only. Notice that the $S^2$ shrinks at the two endpoints of the one dimensional space, so that $M_3$ has the topology of an $S^3$. 
It is actually a \emph{suspension}, \emph{i.e.} a fibration of an $S^2$ over an interval in the coordinate $r$. Locally, supersymmetry turns out to reduce the 10d field equations to the system of ODE's in \eqref{eq:oder}.

The NS-NS flux reads 
\be
H_{(3)} \, = \, -\left(6e^{-A(r)} \,+\, F_{(0)}\,x(r)\,e^{\Phi_{0}(r)}\right)\,\textrm{vol}_{M_3}
\ee
where $F_{(0)}$ is the Romans' mass, while the R-R two-form flux is given by
\be
F_{(2)} = \frac{1}{16}\,e^{A(r)-\Phi_{0}(r)}\,\sqrt{1-x(r)^2}\,\left(F_{(0)}\,x(r)\,e^{A(r)+\Phi_{0}(r)}\, -\, 4\right)\,\textrm{vol}_{S^2}
\ee

It was subsequently found in \cite{Apruzzi:2015zna} that the solutions are determined by a unique function $\beta(y)$ satisfying a single ODE:
\begin{equation}\label{eq:ode}
	(q^2)'= \frac29 F_0 \ ,\qquad q\equiv -\frac{4y\sqrt{\beta}}{\beta'}\ ,
\end{equation}
together with the following constraints 
\begin{equation}\label{eq:Abeta}
\begin{split}
e^{A(y)}\,= \,\frac23 \left(-\frac{\beta'(y)}{y}\right)^{1/4}&\ ,\qquad e^{\Phi_{0}(y)}\,=\,\frac{(-\beta'(y)/y)^{5/4}}{12\sqrt{4 \beta(y) - y \,\beta'(y)}}\ ,\\
x(y)^2\,=&\,\frac{-y \beta'(y)}{4 \beta(y) - y\, \beta'(y)}\ .
\end{split}
\end{equation}
It was shown in \cite{Apruzzi:2015zna} that\eqref{eq:ode} and \eqref{eq:Abeta} together   are equivalent to the system of ODE's \eqref{eq:oder}. The global AdS$_7$ solutions of \cite{Apruzzi:2013yva} may be rewritten in terms of a unique function \cite{Apruzzi:2015zna} $\beta(y)$, which is given by
\be \label{eq:beta}
\beta(y) \ = \ \frac{8}{F_{(0)}} \, \left(y\,-\,y_{0}\right)\,\left(y\,+\,2y_{0}\right)^{2} \ ,
\ee 
where $y_{0}=-\frac{3k_{\overline{D6}}^2}{8F_0}$. \eqref{eq:beta} is a particular solution, with D6 branes at one pole, whereas the other pole is regular. The $\beta$'s which resembles D6, $\overline{{\rm D6}}$ branes and O6 planes singularities are given in \cite{Apruzzi:2015zna}, and nicely summarized in \cite{Apruzzi:2015wna}. 

The coordinate $y$ is defined by the following differential relation 
\be \label{eq:ydef}
dr \,= \,\left(\frac{3}{4}\right)^2\, \frac{e^{3A(y)}}{\sqrt{\beta(y)}}\, dy \ .
\ee

As explained in \cite{Apruzzi:2013yva} and \cite{Gaiotto:2014lca}, there are no regular solutions without any D-brane sources. The only sources that can preserve R-symmetry are D6's, $\overline{\rm D6}$'s
 or O6's, at the poles --- and/or D8's wrapping an $S^2$ at a certain value $r_{\rm D8}$ of $r$. 
When no D8's are present, it is possible to have solutions with an arbitrary number $k_{\rm D6}$ of D6's at the south pole and $k_{\rm \overline{D6}}$ of ${\rm \overline{D6}}$'s at the north pole; 
these solutions have Romans' mass $F_{(0)}$ and NS-NS three-form $H_{(3)}$, restricted by the BI to satisfy 
\be
k_{\rm D6} \,-\, k_{\rm \overline{D6}} \, = \, F_{(0)}\, N \ ,
\ee
where $N$ is the integrated and properly normalized $H$-flux integer. The case when $F_{(0)}\,=\,0$ is just the reduction along the Hopf fiber of the famous AdS$_7\times S^4/\mathbb{Z}_k$ solution.

Studying the asymptotic behavior for the system (\ref{eq:oder}) automatically informs us about the presence of D6, $\overline{\textrm{D6}}$ or O6 sources. 
For example, if we consider $F_0>0$, it follows from (\ref{eq:oder}) that at $x=1$ (the north pole) $A+\Phi$ can go to either $\pm\infty$. For $A+\Phi \to-\infty$ we have that 
\begin{equation}
e^{A(r)} \, \sim \, r^{1/3} \ ,\qquad e^{\Phi(r)} \, \sim \, r\ ,\qquad x(r) \, \sim \, 1 + r^{4/3} \ ,
\end{equation}
which is the correct asymptotic behavior for an $\overline{\textrm{D6}}$ stack. This exercise can be repeated also for D6 branes at the south pole or O6 planes at the north pole, where the positivity of the charges depends on the sign of the function $x$. At the moment the presence of $\overline{\textrm{D6}}$ branes has nothing to do with susy breaking, indeed here we are in curved space and the Killing spinor conditions are modified in a way that does not break supersymmetry.

More solutions can be obtained by introducing D8 branes. In this case, the Romans' mass $F_{(0)}$ jumps because of the presence of the D8's, and correspondingly the metric is continuous but has a discontinuous first derivative. There is an infinite set of solutions \cite{Apruzzi:2013yva,Apruzzi:2015zna}, which are conjectured \cite{Gaiotto:2014lca} to be in one-to-one correspondence with the NS5--D6--D8 system described in section~\ref{sec:braneeng}.

\newsubsubsection{Non-supersymmetric $\textrm{AdS}_7$ counterparts} \subsubseclabel{sec:nonsusyads}
\noindent The non-supersymmetric solutions are in a way very similar to their supersymmetric counterparts. The 10d solution in this case is obtained by setting $X\,=\,2^{-1/5}$ in the \emph{Ansatz} \eqref{eq:DWmetric}.
This gives
\begin{equation}
ds^2_{10}\,= \,\frac{3}{2 \sqrt{2}}\,e^{2A(r)}\, ds^2_{{\rm AdS}_7} \,+\, \frac{1}{ \sqrt{2}}\,ds^2_{M_3}\ ,
\end{equation}
where the metric for $M_3$ is 
\begin{equation}
ds^2_{M_3}\,=\,\left(dr^2\,+\,\frac{e^{2 A(r)} \,\left(1-x(r)^2\right)}{8 \, (1+x(r)^2)}\, ds^2_{S^2}\right)\ .
\end{equation}
The dilaton reads
\begin{equation}
\Phi\,=\,\Phi_{0}(r) \,+\, \frac{1}{2}\, {\rm log}\left(\frac{\sqrt{2}}{1+x(r)^2}\right)\ ,
\end{equation}
where we set $\kappa^3\,=\,8\sqrt{2}$. The $H_{(3)}$ flux is given by
\begin{equation}
H_{(3)}\,=\,-\frac{(e^{-A(r)}\,+\,F_{(0)} \,x(r)\, e^{\Phi_{0}(r)})}{1+x(r)^2}\,{\rm vol}_{M_3}\ .
\end{equation}
Finally $F_{(2)}$ reads
\begin{equation}
F_{(2)}\,=\, \left( \frac{F_{(0)} \, e^{2A(r)} \, x(r) \,\sqrt{1-x(r)^2}}{8(1+x(r)^2)} - \frac{1}{4}\sqrt{1-x(r)^2}\, e^{A(r)-\Phi_{0}(r)}\right) \, {\rm vol}_{S^2} \ ,
\end{equation}
where the functions $A$, $\Phi_{0}$ \& $x$ still satisfy the differential conditions \eqref{eq:oder}. 
Besides some numerical scaling factors, the behavior at the poles of the non-supersymmetric solutions are equivalent to the supersymmetric ones, this makes more challenging the interpretation of the susy braking in terms of brane construction at the conformal point. Moreover, we can actually see that for each supersymmetric solution there exists a very similar non-supersymmetric counterpart, even when allowing for jumps of $F_{(0)}$, 
namely D8 branes wrapping $S^2$ of the suspension geometry. 

\newsubsection{Interpolating 10d flow solution}
\subseclabel{sec:10Dflow}
\noindent We now want to abandon the restriction made in \eqref{const_X} with the aim of finding flow solutions in the $z$ coordinate. By plugging the 10d flow \emph{Ansatz} into the equations of motion 
of massive type IIA given in appendix~\ref{App:MIIA}, one finds that they are implied by the set of ODE's \eqref{eq:oder}, together with the following set of ODE's
\be\label{DWflow_eqn}
\left\{
\begin{array}{lclc}
a'(z) & = & \widetilde{W}(X) & , \\[2mm]
X'(z) & = & -\dfrac{d}{dX} \, \widetilde{W}(X) & ,
\end{array}\right.
\ee 
where the function $\widetilde{W}$ is characterized by the condition
\be
\begin{array}{lcl}
\dfrac{\kappa ^2 }{40 X^8} \, \left(8 X^{10}+8 X^5-1\right) & \overset{!}{=} & \dfrac{1}{2}\,X^{2}\, \left(\dfrac{d\widetilde{W}(X)}{dX}\right)^{2} \, - \, 3 \, \widetilde{W}(X)^{2} \ .
\end{array}
\ee

The equations in \eqref{DWflow_eqn} are first-order conditions for the functions $a(z)$ \& $X(z)$ which, therefore, determine an \emph{extremal} flow. Such a flow solution must correspond to the metric of a non-BPS NS5--D6--$\overline{\textrm{D6}}$ brane system.
The near-horizon limit thereof can be studied by performing a coordinate redifinition that non-trivially mixes $z$ \& $r$, in analogy with the BPS case \cite{NearHorizon}. What is reasonable to expect here in the near-horizon limit is the emergence of a non-BPS NS5 brane. 
This would be in line with \cite{Sen:1999mg}, where non-BPS $(p-1)$ branes were proven to effectively describe a D$p$--$\overline{\textrm{D}p}$ system upon tachyon condesation, this would be consistent with our proposal in section \ref{sec:tachy}.

\newsubsection{7d effective description}
\subseclabel{sec:7dGravity}
\noindent In the previous part of this section we have presented 10d AdS$_7$ solutions which are the candidate gravity duals of a class of interesting novel 6d CFT's. 
In order to be able to holographically derive a few relevant features of the aforementioned CFT's, it turns out to be particularly useful to embed the gravity solution within half-maximal gauged supergravities in $d=7$.

The theory in its ``minimal'' incarnation, \emph{i.e.} restricted to the gravity multiplet, only contains one scalar field and still already captures the essential features of both the supersymmetric and
non-supersymmetric AdS solutions. However, this can be regarded as a truncation of a larger theory containing three vector multiplets and the link to this will be essential in order to capture the full information concerning the spectrum of type IIA closed string 
excitations. We refer to appendix~\ref{App:Gauged_Sugra} for the details of these gauged supergravity theories and their embedding tensor formulations.

\newsubsubsection{AdS vacua and interpolating DW solution}
\subseclabel{subsec:AdS/DW}
\noindent In appendix~\ref{App:Gauged_Sugra} we argue that massive type IIA backgrounds of the form AdS$_{7}\times S^{3}$ can be studied within the $\textrm{ISO}(3)$-gauged supergravity theory in $d=7$ parametrized by the 
embedding tensor in table~\ref{table:ET/fluxes}. In order to study maximally symmetric vacuum solutions, we just need to analyze the critical points of the effective scalar potential (\ref{VHalf_Max})
specified to this case. Furthermore, supersymmetry preserving vacuum solutions can be in this case easily identified by evaluating the gradient of the superpotential (\ref{Wexpr}) at the critical point and checking whether it 
vanishes.

The scalar potential for the relevant truncation \eqref{truncation} then reads
\be\label{V_ISO3}
V_{(\textrm{AdS}_{7}\times S^{3})} \ = \ \frac{1}{2} \, e^{-4 \sqrt{\frac{2}{5}}\,\phi} \, \left(4\theta ^2+e^{\sqrt{10}\,\phi} \left(\tilde{q}^2-3 q^2\right)-4 \theta \, e^{\sqrt{\frac{5}{2}}\,\phi} \,(3q-\tilde{q})\right) \ ,
\ee
where the constants $\theta$, $q$ \& $\tilde{q}$ are respectively related to NS-NS flux, $S^{3}$ curvature and Romans' mass, according to the dictionary in table~\ref{table:ET/fluxes}. Moreover, the scalar $\phi$ in \eqref{V_ISO3} is exactly related to $X$ appearing in section~\ref{sec:10Dgravity} through 
$X \, \equiv \, e^{\frac{\phi}{\sqrt{10}}}$, and it should not to be confused with the scalar in the tensor multiplet of section \ref{sec:6dtheory}.
The above scalar potential can be obtained from the following \emph{superpotential}
\be\label{W_ISO3}
W_{(\textrm{AdS}_{7}\times S^{3})} \ = \ 2 \theta \,  e^{-2 \sqrt{\frac{2}{5}}\, \phi}\,+\,e^{\frac{\phi}{\sqrt{10}}} (3 q-\tilde{q}) \ .
\ee

By minimizing the scalar potential \eqref{V_ISO3}, we find three types of critical points which are collected in table~\ref{table:AdS7}. This set turns out to be included in the exhaustive classification
carried out by \cite{Dibitetto:2015bia} by using the so-called ``going-to-the-origin'' (GTTO) approach \cite{Dibitetto:2011gm}.
In this particular case, there appear three inequivalent AdS critical points, one of which is supersymmetric (hence stable), whereas out of the other two non-supersymmetric ones only one
appears to be perturbatively stable. 
\begin{table}[h!]
\begin{center}
\scalebox{0.94}[0.94]{
\begin{tabular}{| c || c | c | c || c || c | c | c | c |}
\hline
\textrm{ID} & $\theta$  & $q$ & $\tilde{q}$ & $\phi_{0}$ & $V_{0}$ & mass spectrum & SUSY & Stability \\[1mm]
\hline \hline
1 & $\frac{\lambda}{4}$ & $\lambda$ & $\lambda$ & $0$ & $-\frac{15}{8}\,\lambda^{2}$ &
$\begin{array}{cc}0 & (\times\,3)\\[1mm] -\frac{8}{15} & (\times\,1)\\[1mm] 
\frac{16}{15} & (\times\,5)\\[1mm] \frac{8}{3} & (\times\,1) \end{array}$ & \checkmark & \checkmark \\[1mm]
\hline
2 & $\frac{\lambda}{14}$ & $\lambda$ & $-\frac{8}{7}\lambda$ & $0$ & $-\frac{10}{7}\,\lambda^{2}$ & 
$\begin{array}{cc}0 & (\times\,3)\\[1mm] \frac{12}{5} & (\times\,5)\\[1mm] 
\frac{2}{35}\,\left(22\,\pm\,\sqrt{1954}\right) & (\times\,1)\end{array}$ & $\times$ & $\times$ \\[1mm]
\hline
3 & $\frac{\lambda}{4}$ & $\lambda$ & $\lambda$ & $-\sqrt{\frac{2}{5}}\,\log 2$ & $-\frac{5}{2^{7/5}}\,\lambda^{2}$ &
$\begin{array}{cc}0 & (\times\,8)\\[1mm] \frac{4}{5} & (\times\,1)\\[1mm] 
\frac{12}{5} & (\times\,1)\end{array}$ & $\times$ & \checkmark \\[1mm]
\hline
\end{tabular}
}
\end{center}
\caption{{\it All the AdS solutions of half-maximal supergravity in $d=7$ admitting massive type IIA on AdS$_{7}\times S^{3}$ as 10d interpretation. 
Sol.~1 is supersymmetric, whereas 2 \& 3 are non-supersymmetric. Sol.~2 even violates the BF in \eqref{BF}, thus being unstable.} 
\label{table:AdS7}}
\end{table}

We just remind the reader that such vacua will be given by the extrema $P_{0}$ of the potential, while the perturbative stability of the corresponding solutions can be checked by 
computing the eigenvalues of the following mass matrix
\be
\label{mass_matrix}
{\left(m^{2}\right)^{I}}_{J} \ \equiv \ K^{IK} \, \left. \partial_{K}\partial_{J}V\right|_{P_{0}} \ ,
\ee
$K^{IJ}$ being the components of the inverse kinetic metric. In the case of an AdS extremum, each of the eigenvalues of (\ref{mass_matrix}) normalized to the cosmological constant is required to be above the so-called Breitenlohner-Freedman (BF) bound \cite{Breitenlohner:1982jf} 
in order for it to be perturbatively stable. Such a critical value for arbitrary $d$ is given by
\be
\label{BF}
\frac{m^{2}}{|\Lambda|} \ \overset{!}{\geq} \ - \, \frac{d-1}{2\,(d-2)} \ ,
\ee
which equals $-\frac{3}{5}$ in 7d.

Due to our holographic purposes, we can from now on, disregard Sol.~2 since it contains an instability. By taking a closer look at it, one realizes that it has a negative sign of the $F_{(0)}H_{(3)}$ tadpole, 
as opposed to the other solutions. This signals the presence of exceeding $\overline{\textrm{D6}}$ branes which might explain the arising tachyon. 

Focusing on Sol.~1 \& 3, they may be viewed as two different extrema of the same scalar potential by choosing $4\,\theta\,=\,q\,=\,\tilde{q}\,=\,\lambda$. The supersymmetric extremum has been found \cite{Gaiotto:2014lca} to provide the gravity dual of a 6d $(1,0)$ theory, whereas the non-supersymmetric one 
provide a concrete evidence for the existence of a non-supersymmetric 6d CFT. 

A crucial part of our argument is based on the existence of an interpolating static DW solution connecting the two AdS vacua. This solution was originally found in \cite{Campos:2000yu} and we will review
it here. Not only, will it give holographic evidence for an RG flow between the $(1,0)$ and this novel theory, but it will furthermore allow us to argue in favor of the non-perturbative stability
of the non-supersymmetric AdS extremum, thus rendering our statement much more robust.

Since the aforementioned critical points only differ by the position of the $\mathbb{R}^{+}$ scalar $\phi$, the interpolating DW solution can be found within the truncation \eqref{truncation} obtained by only retaining the field content of the gravity mutiplet, \emph{i.e.} the metric, the two-form and $\phi$. 
By then casting the following \emph{Ansatz}
\be
\left\{
\begin{array}{lclc}
ds_{7}^{2} & = & dz^{2} \ + \ e^{2\,a(z)} \, ds_{\textrm{Mkw}_{6}}^{2} & , \\[2mm]
\phi & = & \phi(z) & ,
\end{array}
\right.
\ee
the DW equations equations read
\be
\left\{
\begin{array}{lclc}
15 \, \left(a^{\prime}\right)^{2} \, - \, \frac{1}{4} \, \left(\phi^{\prime}\right)^{2} \, + \, V & = & 0 & , \\[2mm]
\phi^{\prime\prime} \, + \, 6 \, a^{\prime}\,\phi^{\prime} \, - \, 2 \, \partial_{\phi}V & = & 0 & ,
\end{array}
\right.
\ee
where $^{\prime}$ denotes a derivative w.r.t. the $z$ coordinate.
The interpolating solution that we are looking for describes the flow between a supersymmetric and a non-supersymmetric critical point and therefore, it 
is non-BPS. However, by making use of the Hamilton-Jacobi (HJ) formalism, one can rewrite the above second-order problem as the following system of first-order flow equations \cite{Campos:2000yu}
\be
\label{1storder}
\left\{
\begin{array}{lclc}
a^{\prime} & = & \frac{1}{5\sqrt{2}} \, \widetilde{W} & , \\[2mm]
\phi^{\prime} & = & -\sqrt{2} \, \partial_{\phi}\widetilde{W} & ,
\end{array}
\right.
\ee
provided that the scalar potential $V$ can be rewritten by using $\widetilde{W}$ as
\begin{equation}
\label{VfromfakeW}
\begin{array}{lcl}
V & = & \frac{1}{2}\, \left(\partial_{\phi}\widetilde{W}\right)^{2} \, - \, \frac{3}{10} \, \widetilde{W}^{2} \ .
\end{array}
\end{equation}
Note that the above condition defines the HJ generating function $\widetilde{W}$ as a \emph{fake superpotential}. Although such a function has in general
nothing to do with supersymmetry, the (\ref{VfromfakeW}) always admits the actual superpotential (\ref{Wexpr}) as a solution.
Moreover, when choosing $\widetilde{W} \, = \, W$, the first-order equations in (\ref{1storder}) precisely reduce to BPS conditions for a supersymmetric
DW.

\begin{figure}[t!]
\begin{center}
\includegraphics[scale=1.1]{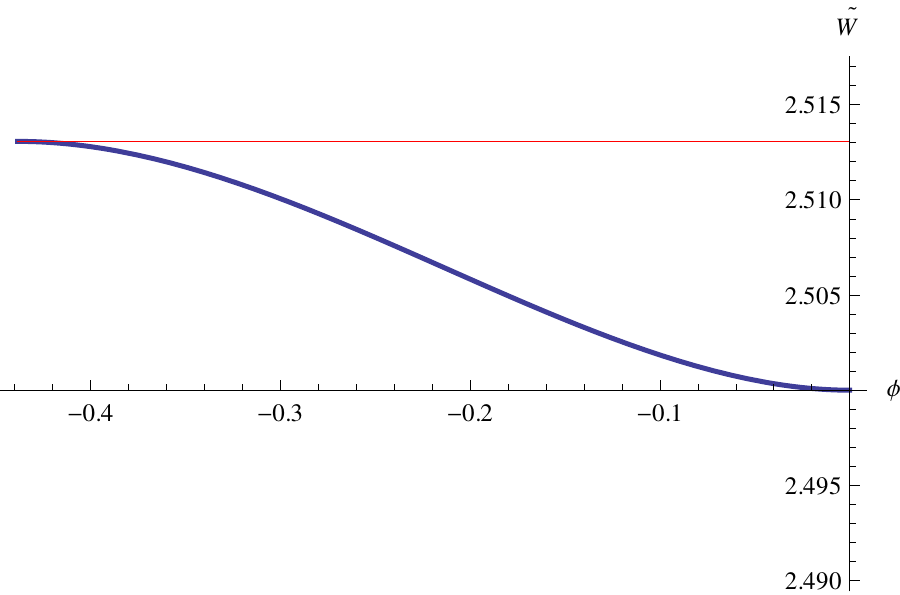}
\end{center}
\caption{{\it The profile of the fake superpotential $\widetilde{W}(\phi)$ solving the differential condition (\ref{VfromfakeW}) on the interval $\left[-\sqrt{\frac{2}{5}}\,\log 2,\,0\right]$.
Note that, in contrast with the actual superpotential obtained by specifying (\ref{Wexpr}) to this gauging, $\widetilde{W}$ has a stationary point at both extrema, \emph{i.e.} both Sol.~1 \& 3 are
fake-supersymmetric.}}
\label{fig:fake_W}
\end{figure}

\begin{figure}[t!]
\begin{center}
\includegraphics[scale=1.1]{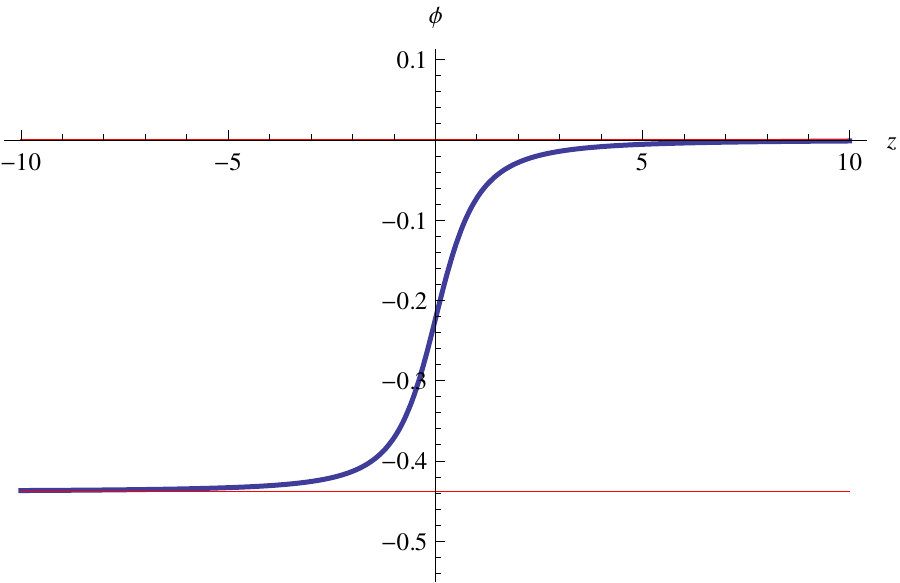}
\end{center}
\caption{{\it The profile of the scalar field $\phi(z)$ along the non-BPS DW solution. Note that it correctly interpolates between the two AdS critical points labeled by 1 \& 3 in table~\ref{table:AdS7}.
In such limits the scale factor $a(z)$ correctly asymptotes to $\frac{z}{\ell}$, where $\ell \, = \, \sqrt{-\frac{V_{1}}{15}}$ or $\sqrt{-\frac{V_{3}}{15}}$, respectively.}}
\label{fig:DWsol}
\end{figure}

Interestingly, in our case the (\ref{VfromfakeW}) happens to have two inequivalent global solutions, the first one being the actual superpotential and 
the second one being a fake superpotential which can only be determined numerically. The corresponding function is plotted in figure~\ref{fig:fake_W}. 
This allows one to solve the flow equations in (\ref{1storder}). The $z$ profile of the DW is shown in figure~\ref{fig:DWsol}.
\newpage
\vspace{10cm}
\newpage
\newsection{Holographic Description}
\seclabel{sec:CFT6}
\noindent In the previous section we presented a DW solution interpolating between the supersymmetric and non-supersymmetric AdS vacua which suggested the existence of an RG flow between a 6d $(1,0)$ theory in the UV and a novel local 6d CFT in the IR.

The aim of this section is to examine all the properties of the theory at the IR fixed point by means of holographic techniques. This analysis includes a calculation of the conformal anomaly and a validity check of the ``a-theorem" throughout the flow. We observe that both tests give interesting results about the novel CFT$_6$. Moreover, we calculate the spectrum of conformal dimensions at both fixed points. Together with the previous analysis, this gives a further evidence about the existence of an interacting CFT$_6$ in the IR.
\newsubsection{Conformal anomaly}
\noindent  In general, there is a deep intuition that computing the conformal anomaly for a CFT is a way to measure its degrees of freedom. In this way, it is very important to test this quantity also in our 6d CFT in order to understand what the expected properties of the theory are. The universal trace anomaly in even spacetime dimensions $d$ was computed holographically in the seminal paper \cite{Henningson:1998gx}, this yields
\be\label{Trace_anomaly}
\langle {T^{\mu}}_{\mu} \rangle_{d} \, = \, \frac{\ell^{d-1}}{G_{\rm N}} \, a_{(d)} \ ,
\ee
where $a_{(d)}$ denotes a suitable polynomial containing degree $d/2$ curvature invariants in the components of the Riemann tensor. In particular, the explicit general form of the aforementioned anomaly
polynomial implies $a\,=\,c$ for all CFT's admitting a holographic dual. In order to infer the correct behavior of the $a$ coefficient, we follow the procedure outlined in \cite{Gaiotto:2014lca,Cremonesi:2015bld}, consisting in computing the following integral
\be
\mathcal{F} \, = \, \int e^{5A-2\Phi}\,\textrm{vol}_{3} \ .
\ee
It can be easily seen that the resulting $N^{3}$ behavior of the free energy obtained in \cite{Gaiotto:2014lca} for the supersymmetric case will still hold in the non-supersymmetric setting analyzed here. 
The intuitive reason for this is that the above integral only differs by certain numerical factors related to powers of the scalar $X$\,\,\footnote{The ratio of the free energies for the susy and non-susy solution is $
\frac{\mathcal F_{{\rm susy}}}{\mathcal F_{\rm non-susy}}=X^{20}$.
}.
Furthermore, this novel example of conformal fixed point gives us the opportumity of testing the validity of an a-theorem within a non-supersymmetric context in 6d, \emph{i.e.}
\be
a_{\rm UV} \, > \, a_{\rm IR} \,\,\, . \nonumber
\ee
The above statement is known to hold for SCFT$_6$'s \cite{Cordova:2015fha}, although a general proof outside of the supersymmetric case is still lacking.
To verify this in our case, we use \eqref{Trace_anomaly} to conclude that 
\be
\begin{array}{lclclclc}
a & \sim & \ell^{5} & \sim & V_{\rm crit.}^{-\frac{5}{2}} & \sim & \widetilde{W}_{\rm crit.}^{-5} & ,
\end{array}
\ee
where, in the last step, we have used that $\partial_{\phi}\widetilde{W}=0$ at both ends of the flow. By taking a look at figure~\ref{fig:fake_W}, this implies that $c$ decreases monotonically along the flow confirming the expectation from the a-theorem.  Notice that this analysis is equivalent to that of \cite{Girardello:1998pd}, provided that we identify the central charge function as $c(z) \, \sim \, \left(T_{zz}\right)^{-\frac{5}{2}}$.

\newsubsection{Operator spectrum}
\noindent The upshot of the previous section is that our novel CFT$_6$ has a positive conformal anomaly and seems to satisfy our ``na\"ive" expectations from the a-theorem. Now,  given that we are describing this theory as a new fixed point associated to an RG flow, it is interesting to compute the operator spectrum of the theory both at the supersymmetric (UV) and non-supersymmetric (IR) point and to understand which operator is responsible for triggering the flow. 

We can address this problem in the framework of the effective 7d description of gauged supergravity as introduced in appendix~\ref{App:Gauged_Sugra}. The masses associated with the $\textrm{AdS}_7$ states both in the supersymmetric and non-supersymmetric case are the following:\footnote{Note that the values appearing here are related to those in table~\ref{table:AdS7} upon using $\Lambda\,=\,-\frac{15}{\ell^2}$.} 
\be\label{eq:Adsmasses}
\begin{array}{lcclc}
\textrm{SUSY} : &  m_{i}^2\,\ell^2  & = & \{0\,(\times 3),\quad -8,\quad 16\,(\times 5),\quad 40\} & , \\[4mm]
\textrm{NON-SUSY} : &  m_{i}^2\,\ell^2 & = & \{0\,(\times 8), \quad 12, \quad 36\} & .
\end{array}
\ee
At this point, we can use a very well known formula which relates the value of the supergravity masses to the conformal dimension of the operators in the dual CFT. Namely, in 6d we have
\be\label{eq:conformaldim}
\Delta_{i}\,\left(\Delta_{i}\,-\,6\right) \, = \, m_{i}^{2}\,\ell^{2} \ ,
\ee
where $\Delta_{i}$ denotes the conformal dimension of the operator $\mathcal{O}_{i}$ in the dual CFT.

Thanks to \eqref{eq:conformaldim}, we can construct the spectrum of operators at a given conformal fixed point. Moreover, it turns out to be possible to keep track of the conformal dimensions of all operators
throughout the whole flow. This can be done holographically by exploiting the information encoded within the scalar sector of the 7d effective supergravity introduced in appenidx~\ref{App:Gauged_Sugra}.

The DW solution connecting our AdS vacua preserves a residual $\textrm{SO}(3)$ symmetry. In particular, in the supersymmetric extremum, such a symmetry can be interpreted as 
the $\textrm{SU}(2)_{R}$ R-symmetry group of the dual $(1,0)$ SCFT. Notice that the non-supersymmetric extremum will also share the same residual symmetry, this is interpreted as a global symmetry of the new theory and it is no longer related to R-symmetry in any way.
This implies that the scalar mass spectrum of both solutions be organized into $\textrm{SO}(3)$ irrep's.\footnote{In order to check this, we perform the following branching
\be
\begin{array}{cclc}
\mathbb{R}^{+}_{\phi} \, \times \, \textrm{SL}(4,\mathbb{R}) & \supset & \textrm{SO}(3) \\[4mm]
\textbf{1}_{(0)} \, \oplus\, \textbf{15}_{(0)} & \longrightarrow & \textbf{1} \, \oplus\, \textbf{1} \, \oplus\, \textbf{3} \, \oplus\, \textbf{5} \, \oplus\, \textrm{unphys.} & , 
\end{array} \nonumber
\ee
where the label ``unphys.'' denotes all irrep's corresponding to compact $\textrm{SL}(4,\mathbb{R})$ generators, which do not correspond to physical scalars in that they can be gauged away. }

Notice that the spectra in \eqref{eq:Adsmasses} are indeed organized into the expected irrep's of $\textrm{SO}(3)$. Furthermore, by combining \eqref{eq:conformaldim} with the analysis of the eigenvectors
of the mass matrix at the critical points, the evolution of the operator spectrum along the RG flow can be obtained unambiguously. The corresponding results are collected in table~\ref{table:Operators}.
\begin{table}[h!]
\renewcommand{\arraystretch}{1}
\begin{center}
\scalebox{1}[1]{
\begin{tabular}{ c || c | c }
$\textrm{SO}(3)$ irrep's of $\varphi^{i}$ & UV $\Delta_{i}$ & IR $\Delta_{i}$ \\
\hline 
$\mathbf{1}$ & 4 &\,3 + $\sqrt{21}$\,$\approx 7.58$  \\
$\mathbf{1}$ & 10 &\,$3 \left(1+\sqrt{5}\right)$\,$\approx 9.71$\\
$\mathbf{3}$ & 6 & 6\\
$\mathbf{5}$ & 8 & 6
\end{tabular}
}
\end{center}
\caption{{\it The conformal dimension of the operators in the dual conformal fixed points at both ends of the RG flow, \emph{i.e.} the supersymmetric one in the UV and the non-supersymmetric one in the 
IR. $\varphi^i$ indicates the set of scalars of the 7d Supergravity.} 
\label{table:Operators}}
\end{table}

The physical interpretation of the operators appearing in table~\ref{table:Operators} is the following. Due to the fact that the DW solution discussed here may be found within the truncation \eqref{truncation},
the only scalar assuming a dynamical profile is $\phi$. As a consequence, the only operator acquiring a vev during the flow must correspond with the first singlet in table~\ref{table:Operators} and hence
it is the one responsible for triggering the RG flow in the CFT. This operator has dimension $4$, \emph{i.e.} it is a relevant operator in 6d and it is interpreted as a mass deformation. As one can easily
see, the remaining operators listed above are irrelevant or at most marginal and as such, do not play any role in this context.

Let us denote by $\mathcal{O}_{X}$ the operator of dimension 4. The analytical expression of its conformal dimension $\Delta_{X}$ may be determined all along the flow by computing the mass of the dual 
canonically normalized scalar field $\phi$. This reads
\be
m^{2}_{\phi} \, = \, -\frac{3}{5}\,\left(\partial_{\phi}\widetilde{W}\right)^{2} \, - \, \frac{3}{5}\,\widetilde{W}\,\partial_{\phi}^{2}\widetilde{W} \,+\, \left(\partial_{\phi}^{2}\widetilde{W}\right)^{2} \, + \, \partial_{\phi}\widetilde{W}\,\partial_{\phi}^{3}\widetilde{W} \ ,
\ee
where $\widetilde{W}(\phi)$ is the fake superpotential plotted in figure~\ref{fig:fake_W}. Note that the above quantity stays meaningful throughout the flow since $\phi$ is an eigendirection of the mass 
matrix and does not mix to any other scalars. Finally, by using \eqref{eq:conformaldim}, we computed the corresponding conformal dimension $\Delta_{X}$ and found that it has a monotonic behavior starting from $4$ in the UV
and ending up at $3+\sqrt{21}$ in the IR.

Our choice suggests a correspondence with the dual scalar $X(z)$ responsible for the interpolating solution in supergravity. In section \ref{sec:tachy} we proposed a stringy picture of the RG flow. The  presence of anti-branes hints to a move in the tensor branch that culminates in a new non-supersymmetric fixed point. As explained, in the limit where the distance between NS5 branes goes to zero the tachyons do not couple with any fields in the resulting theory. This occurrence can be also sketched within a 6d effective action description.\\
Indeed, because of symmetry arguments we can single out the following  6d effective Lagrangian \footnote{Here and in what follows in the rest of this section, $\phi$ denotes again the scalar of the tensor multiplet.}
\be
\mathcal{L}_{\rm eff} \, = \, \phi\,\textrm{Tr}(F^2) \, + \, (\partial \phi)^2 \, + \, \phi\,\mathcal{O}_{X} \, + \,  ...\ ,
\ee
where the remaining terms are not important for the current discussion. The coupling $\phi\,\mathcal{O}_{X} $ is invariant under $\textrm{SU}(2)_R$ and matches the description with the potential introduced in section \ref{sec:tachy}. In order to get rid of the above coupling we need to go at the point $\phi = 0$, \emph{i.e} at the collision of the NS5's. However, notice that in such a limit the effective action approach loses its meaning and we see the emergence of a conformal fixed point. Another possibility is that $\mathcal{O}_{X} \sim \phi^2$ and the term in the effective action triggering the flow along the tensor branch is a mass term of the form: $m\phi^2$. Because of the evidence coming from the holographic solution, we conclude that, in our setup, the susy breaking deformation involves a coupling with the tensor multiplet scalar.

It is worth mentioning that the appearence of operators in the IR with non-integer conformal dimensions should not be regarded as surprising but rather as a signal of an ``interesting" interacting theory. The reason for this goes as follows. Any free CFT admits a Lagrangian description. If that is the case the conformal dimensions of primary operators are integers or semi-integers. In this way the OPE can only be trivial because all the 3-point functions are trivial, it follows that a free CFT cannot have ``new" operators with non-(semi)integers dimensions.
 
Moreover, such a situation should be generically expected whenever considering non-supersymmetric flows where all operators remain unprotected and tend to develop non-vanishing anomalous dimensions.
In this sense, the dual effective gauged supergravity approach turns out to be rather fruitful because it gives us an opportunity to compute such anomalous dimensions exactly.
\newpage

\newsection{Stability checks and evidence for anti-branes susy breaking in 6d theories}\seclabel{sec:stability}
\noindent In this section we perform some checks regarding various stability issues of the gravitational solutions in order to provide further confirmation to our holographic picture. 

Before moving to the stability analysis of the non-supersymmetric vacuum at a non-perturbative level, let us first add an important remark concerning its perturbative stability. Within the context of the effective 7d supergravity theory coupled to
3 vector multiplets, we have essentially computed the mass spectrum of the closed string zero modes. Going beyond the zero modes, the spin-2 spectrum was recently analyzed in \cite{Passias:2016fkm} and it 
was found to be positive.

Despite that, one may still worry about
the size of perturbative corrections both in $\alpha^{\prime}$ and $g_{s}$. However, these stringy corrections will come with powers of $\frac{1}{N}$, $\frac{1}{k}$ or $\frac{1}{F_0}$, and in the holographic limit where the flux integers are large (which also means large size of the internal $S^{3}$ when expressed in string units), we still have some control on them even in the non-supersymmetric case. In fact, there will be a regime where the supergravity description is still good enough. More precisely, since there is a finite distance between the masses of all the closed string modes and the BF bound, we can conclude that there exist big enough flux integers such that all the masses remain above the BF bound, even when we include these perturbative string corrections. In other words, when any modes do not saturate the BF bound, the stability cannot be violated in the limit where the stringy perturbation are arbitrarily small, which is exactly the holographic limit. Similar arguments where used in \cite{Gaiotto:2009mv,Narayan:2010em}.

\newsubsection{Non-perturbative stability}
\subseclabel{subsec:stability}
\noindent In order to support our holographic evidence of non-supersymmetric 6d CFT's, we address here the issue of non-perturbative stability of Sol.~3 in table~\ref{table:AdS7}. Generically, perturbatively stable
 AdS critical points are not protected from quantum tunneling effects that might result in a decay towards the true vacuum of the theory. Such an effect may occur through spontaneous bubble nucleation
processes (see for example \cite{Witten:1981gj,Horowitz:2007pr}) which are described by gravitational instanton solutions. 

In \cite{Coleman:1980aw,Brown:1988kg} the above situation was considered within a theory of gravity, where the true and false vacuum, respectively characterized by $\Lambda_{1} \, < \, \Lambda_{2}$, are separated by a DW. Gravitational instantons were found to provide a finite contribution to the Euclidean action provided that the following condition be satisfied
\begin{equation}
\label{BTbound}
\tau_{\textrm{DW}} \ < \ \frac{2}{\sqrt{3}}\,\left(\sqrt{\left|\Lambda_{1}\right|}\,-\,\sqrt{\left|\Lambda_{2}\right|}\right) \ ,
\end{equation}
where $\tau_{\textrm{DW}}$ denotes the tension of the separating wall.

In our case, the supersymmetric and the non-supersymmetric AdS vacua are separated by a static extremal DW, which then happens to saturate the CDL bound \cite{Cvetic:1992bf}. This may be already regarded
as a sign of a decay channel that does not yield any finite contribution to the Euclidean action. Contrary to static DW, any other time dependent interpolating solution will necessarily give rise to a finite contribution to the Euclidean action thus resulting in a non-perturbative instability.

Supersymmetric critical points have been proven not to be affected by such decay processes \cite{Witten:1982df,Weinberg:1982id} due to the existence of positive energy theorems, whereas
the issue might become very subtle when it comes to non-supersymmetric solutions. However, in \cite{Boucher:1984yx} a generalization of the above positive energy theorems was developed and a concrete criterion for stability was established. 
The argument makes use of the equivalence between fake supersymmetry (the existence of a fake superpotential) and HJ flows.
This issue has been recently addressed more in detail and in a more general context in \cite{Danielsson:2016rmq}, where the global properties of fake superpotentials are found to be crucial for
non-perturbative stability. In particular, the existence of an interpolating static domain wall turns out to be a sufficient condition for non-perturbative stability. 

Adopting this strategy in our case, the above criterion boils down to checking the existence of a suitable \emph{globally defined} function (\emph{i.e.} a fake superpotential, or equivalently a HJ generating functional) 
$f(\phi)$ satisfying $V\, = \, \frac{1}{2}\, \left(\partial_{\phi}f\right)^{2} \, - \, \frac{3}{10} \, f^{2}$ and 
 such that
\be
\begin{array}{lcl}
f'(\phi) = 0 \ , & &  \textrm{at both extrema} \ ,\\[2mm]
V_{f}(\phi)  \, \leq \, V(\phi) \ , & & \forall \ \phi \, \in \ \textrm{decay path}\ ,
\end{array}
\ee
where $V_{f}\, \equiv \, - \, \frac{3}{10} \, f^{2}$ and $V(\phi)$ is given in \eqref{V_ISO3}. The fake superpotential $\widetilde{W}$ plotted in figure~\ref{fig:fake_W}
defining the extremal flow \eqref{1storder} happens to obey the above inequality, thus proving a positive energy theorem for the non-supersymmetric AdS vacuum. This implies the stability of the non-susy AdS critical point. Moreover, arbitrarily small perturbative corrections may shift a little bit the position of the vacua, but also they will change the tension of the Domain Wall by a small amount, hence they will not spoil the fact that the Domain Wall is static.

\newsubsection{Presence of anti-branes: a physical argument}
\subseclabel{subsec:antibranes}
\noindent As we already mentioned the non-supersymmetric AdS$_7$ solutions of section \ref{sec:nonsusyads} are very similar to the supersymmetric one found in \cite{Apruzzi:2013yva,Apruzzi:2015zna} and summarized in section \ref{sec:susyads}, namely they have analogous boundary behaviors close to the poles of the $S^3$. However, something happens in the interior of the $S^3$, and the geometry of the susy solutions gets deformed, namely the profiles of fields in the non-supersymmetric solutions are a bit different from the susy ones. 

In section \ref{sec:10Dgravity}, we conjectured that a 6d susy breaking is realized by inserting the same amount of D6 and $\overline{\textrm{D6}}$ between two NS5's when the theory is resolved in its tensor branch (\emph{i.e.} all the NS5's are separated in the $x^6$ direction, or at least partially). As explained in section \ref{sec:tachy}, the brane system is then unstable and it decays to a point in moduli space where the NS5 branes are all coincident, \emph{i.e.} to a conformal point or to a holographic AdS$_7$ vacuum when taking the near-horizon limit. The susy-breaking $\overline{\textrm{D6}}$'s are now annihilated in the vacuum and therefore we do not observe any significant change at the poles in the branes charges. 

However, we can still study the ratio of the tensions of our AdS$_7$ solutions, comparing it with the following quantity
\begin{equation}
\frac{T_{\rm D6}\,+\,T_{\overline{\rm D6}}}{T_{{\rm D6}, \, {\rm susy \; sol.}}}\ ,
\end{equation}
where $T_{{\rm D6}, \, {\rm susy \; sol.}}$ is the tension of the D6 branes at the poles in a supersymmetric AdS solution, whereas $T_{\rm D6}\,+\,T_{\overline{\rm D6}}$ is the sum of the tension of D6 branes and $\overline{\textrm{D6}}$ branes\footnote{Similar analysis about the tension of D6/$\overline{{\rm D6}}$ systems appeared in \cite{Blaback:2011pn}.}. In general we have that for a bound state of branes 
\begin{equation}
T_{\rm D6}\,+\,T_{\overline{\rm D6}} \ \geq \ T_{{\rm D6}+\overline{\rm D6}}\ \geq \ T_{{\rm non-susy\; sol.}}\ ,
\end{equation} 
where $T_{{\rm non-susy\; sol.}}$ is the tension of the branes at the poles of a non-supersymmetric solution, when everything is already annihilated and tachyons condensed. We now have the following inequality
\begin{equation}
\frac{T_{\rm D6}\,+\,T_{\overline{\rm D6}}}{T_{{\rm D6}, \, {\rm susy \; sol.}}} \,\geq \,\frac{T_{{\rm non-susy\; sol.}}}{T_{{\rm D6},\, {\rm susy \; sol.}}}
\end{equation}
where we used the time-time component of the Einstein tensor (\emph{i.e.} $|R_{tt}\,-\,\frac{1}{2}g_{tt}\, R|$ ) to estimate the tension at both AdS$_7$ points.
By plugging all the quantities in \eqref{eq:ydef} and \eqref{eq:Abeta}, we can extract the leading order coefficient in two variables $(z, y)$, which in terms of the constant $X$ is
\begin{equation}\label{eq:esttens}
T\,\sim\, \frac{X^5}{(-1 + 8 X^5 (1 + X^5))}\ .
\end{equation}

For the susy case, we computed $T_{{\rm D6},\, {\rm susy \; sol.}}$ by plugging in \eqref{eq:esttens} $X=1$, and $T_{{\rm non-susy\; sol.}}$ by plugging in \eqref{eq:esttens} $X^{5}=\frac{1}{2}$, which lead to 

\begin{equation}
\frac{T_{{\rm non-susy\; sol.}}}{T_{{\rm D6},\, {\rm susy \; sol.}}}\, = \,\frac{3}{2} \quad \Longrightarrow\quad  \frac{T_{\rm D6}+T_{\overline{\rm D6}}}{T_{{\rm D6},\, {\rm susy \; sol.}}} \,\geq \,\frac{3}{2}\ .
\end{equation}
First of all, the fact that the tension of the non-supersymmetric solution is bigger than the supersymmetric one reflect the fact that somehow in the non-supersymmetric vacuum there should be a remnant of some extra brane sources. 

We can now give a bound on the number of branes and anti-branes we need to add in order to reach a non-supersymmetric AdS point. The tension can be written as $T_{D6}\sim k |Q_{D6}|$. Moreover charge conservation at the poles implies the following constraint

\begin{equation}
T_{{\rm D6},\, {\rm susy \; sol.}} \,\sim \,(k_{D6}-k_{\overline{D6}})\,Q_{D6}\ .
\end{equation}

We then get a new inequality,
\begin{equation}
\frac{k_{\rm D6}\,+\, k_{\overline{\rm D6}}}{k_{\rm D6}- k_{\overline{\rm D6}}} \,\geq \,\frac{3}{2}\ ,
\end{equation}
which can be expressed as a lower bound for the number of $\overline{\textrm{D6}}$
\begin{equation}
k_{\overline{\rm D6}}\,\geq  \, \biggl[ \frac{k_{\rm D6}}{5} \biggr]\ .
\end{equation}
We have given here an evidence for the presence of $\overline{\textrm{D6}}$'s during the non-supersymmetric DW interpolating flow between the two AdS$_7$ vacua. The D6/$\overline{\textrm{D6}}$ system, that we conjecture to give a holographic non-supersymmetric AdS$_7$ vacuum in the IR. This picture, indeed, seems to be consistent with an estimation of the ratio for the tension of the two AdS solutions. Finally, it would be interesting to understand better this bound on the number of $\overline{\textrm{D6}}$'s. This bound might be due to the effect of the supersymmetric D6's stretching between the two NS5's as  gravitational background for the tachyon condensation given in section \ref{sec:tachy}.
\newpage

\newsection{Conclusions}\seclabel{sec:conclusion}

\noindent In this paper we conjectured the existence of a class of novel six-dimensional and non-supersymmetric conformal field theories. These theories are strongly interacting fixed points obtained by means of a non-supersymmetric RG flow.

We have been able to identify the RG flow as an interpolating DW solution of gauged supergravity. The DW connects two AdS$_7$ vacua, a supersymmetric one in the UV and a non-supersymmetric one in the IR. A careful analysis of the supergravity spectrum has been carried out in order to exclude the presence of tachyons in the non-supersymmetric AdS$_7$ vacua. Using holography, we also calculated the conformal dimensions spectrum of dual theories at both fixed points. The outcome of this calculation is that there is only one operator acquiring a vev during the flow, which has dimension 4, \emph{i.e.} a relevant operator in 6d. It is responsible for triggering an RG flow between dual CFT's. 

As already mentioned, all the solutions of 7d gauged supergravity used in this paper have a known string theory lift. 
Here we proposed a susy-breaking mechanism at the level of massive IIA brane constructions for 6d $(1,0)$ theories. We created a meta-stable non-supersymmetric gauge theory in the tensor branch by means of $\overline{\textrm{D6}}$ branes and we explained how the configuration reaches a stable conformal fixed point in the IR. In particular we outlined a mechanism relating the presence of $\overline{\textrm{D6}}$ branes to a holographic RG flow and we gave some indications about the fate of the open string tachyons along the flow.

The possibility of a non-supersymmetric 6d CFT is quite attractive also from a field theoretical point of view. At the moment, many properties of six-dimensional conformal field theories outside the supersymmetric case are unknown. We attempted to clarify some features of this new example.  More precisely, we computed the holographic anomaly and we tested the validity of the ``a-theorem" along the flow in terms of the gravitational stress energy tensor. Both tests confirm general expectations about 6d CFT's. Of course, much more studies still have to be done to understand this example and we hope that our work is useful in exploring further properties of 6d CFT's.


Finally, we would like to mention some possible generalizations of our analysis. We believe that some features of this work can be extended to other dimensions, as this could deepen the  understanding of non-supersymmetric RG flows in gauged supergravity and their relationship with AdS/CFT. It would also be very interesting to investigate the possibility of new analytic non-supersymmetric solutions as well as interpolating solutions of massive IIA.

The fruitful interaction between anti-branes in AdS and dual CFT observed in this paper is quite remarkable. The emergence of a conformal fixed point resolves in a natural way a number of problematic stability issues. It would be really fascinating to understand if conformal field theory fixed points might play a role also in the stabilization of uplifting anti-branes used in model building scenarios.

From a field theoretical point of view, more work should be done to assess the validity of our susy-breaking mechanism. The absence of a Lagrangian description makes things more complicated and we are not completely certain about the legitimacy of an effective action analysis in this context. A problem that should be immediately addressed concerns the operators' couplings in the effective theory. For instance, there should be a symmetry argument that prevents the coupling of the open string tachyon to other fields in the IR theory as indicated by our stringy construction. In addition, it would be useful to classify all the deformations of a 6d CFT that might lead to the kind of IR fixed point described here. This analysis could also reveal potential further examples of non-supersymmetric CFT$_6$.            

\vspace{2cm}
\noindent {\it Acknowledgements}: It is a pleasure to thank Antonio Amariti, Johan Bl{\aa}b{\"a}ck, Ulf Danielsson, Marco Fazzi, Guido Festuccia, Johnathan Heckman, Pietro Longhi, Noppadol Mekareeya, Anton Nedelin and especially Alessandro Tomasiello for discussions and correspondence related to this work. 
We also thank Antonio Amariti, Fridrik Gautason, Johnathan Heckman, Alessandro Tomasiello, Thomas Van Riet and Marco Zagermann for comments on a draft version of this manuscript.\\
FA thanks the theory groups at Columbia University and
the ITS at the CUNY\ graduate center for hospitality during the completion of this work.
The work of FA is supported by NSF CAREER grant PHY-1452037. FA 
also acknowledge support from the Bahnson Fund at UNC Chapel Hill as well as the
R.~J. Reynolds Industries, Inc. Junior Faculty Development Award from the Office
of the Executive Vice Chancellor and Provost at UNC Chapel Hill.
The work of GD is supported by the Swedish Research Council (VR). The work of LT is supported in part by VR grants \#2011-5079 and \#2014-5517, in
part by the STINT grant and in part by the Knut and Alice Wallenberg Foundation.

\newpage

%
%

\appendix

\renewcommand{\newsection}[1]{
\addtocounter{section}{1} \setcounter{equation}{0}
\setcounter{subsection}{0} \addcontentsline{toc}{section}{\protect
\numberline{\Alph{section}}{{\rm #1}}} \vglue .6cm \pagebreak[3]
\noindent{\bf Appendix {\Alph{section}}:
#1}\nopagebreak[4]\par\vskip .3cm}

\newsection{Massive type IIA supergravity} \seclabel{App:MIIA}
\noindent In this appendix we review our working conventions concerning \textit{massive} type IIA supergravity in ten dimensions. The bosonic part of its string-frame Lagrangian is given by
\be
\label{action_IIA}
S_{\textrm{IIA}} \ = \ \frac{1}{2\kappa_{10}^{2}} \, \displaystyle\int d^{10}x \, \sqrt{-g} \, \left[e^{-2\Phi}\Big(R \, + \, 4  (\partial\Phi)^{2} 
\, - \, \frac{1}{2} |H_{(3)}|^{2} \Big) \, - \, \frac{1}{2} \sum\limits_{p=0,2,4} |F_{(p)}|^{2}\right] \, + \, S_{\textrm{top.}} \ , 
\ee
where $\,|H_{(3)}|^{2} \ \equiv \ \frac{1}{3!} \, H_{(3)MNP}{H_{(3)}}^{MNP}\,$ and $\,|F_{(p)}|^{2} \ \equiv \ \frac{1}{p!} \, F_{(p)M_{1}\dots M_{p}}{F_{(p)}}^{M_{1}\dots M_{p}}$ with $M=0,...,9$. The above action contains a topological term of the form
\be
\begin{array}{lcl}
S_{\textrm{top.}} & = & -\frac{1}{2} \, \displaystyle\int \left(B_{(2)} \wedge dC_{(3)} \wedge dC_{(3)} \, - \, 
\tfrac{1}{3} F_{(0)} \wedge B_{(2)} \wedge B_{(2)} \wedge B_{(2)} \wedge dC_{(3)} \right. \\[2mm]
& + & \left.\frac{1}{20} F_{(0)} \wedge F_{(0)} \wedge B_{(2)} \wedge B_{(2)} \wedge B_{(2)} \wedge B_{(2)} \wedge B_{(2)} \right) \ .
\end{array}
\ee

\noindent We will now review here set of 10d equations of motion (EOM) and Bianchi identities (BI) which follow from the action \eqref{action_IIA}. 
The equations of motion for $B_{(2)}$, $C_{(1)}$ and $C_{(3)}$ are respectively given by
\be \label{eq:fluxeom}
\begin{array}{rclcccc}
d \left(e^{-2 \Phi }\ast_{10} H_{(3)}\right) & = & 0 &  & & & , \\[2mm]
\left(d \,+\, H_{(3)}\wedge \right)(\ast_{10} F_{(p)}) & = & 0 &  & & (p\,=\,2,\,4) & , \\[2mm]
\end{array}
\ee
whereas the one for the 10d dilaton $\Phi$ reads
\be
\label{eq:dileom}
\begin{array}{rrclc}
\square \Phi \, - \, \left|\partial \Phi \right|^2 \, + \, \frac{1}{4} R \, - \, \frac{1}{8}\left|H_{(3)}\right|^2 & = & 0 & ,
\end{array}
\ee
where $R$ is the 10d scalar curvature and $\square$ is the ten-dimensional (curved) Laplacian operator. The 10d Einstein equations take the standard form\footnote{Note that we have set $\kappa_{10}\,=\,1$.}
\be
\label{10s_Einstein}
\begin{array}{rrclc}
R_{MN} \ - \ \frac{1}{2} \, T_{MN} & = & 0 & ,
\end{array}
\ee
where the symmetric energy-momentum tensor $T_{MN}$ is defined as
\begin{eqnarray}
T_{MN} & = & e^{2\Phi} \,\sum\limits_{p}  \left(  \frac{p}{p!} \,  F_{(p) M M_{1}\dots M_{p-1}}  F_{(p)N}^{\phantom{(p)N}M_{1}\dots M_{p-1}} \, - \, \frac{p-1}{8} \,  g_{MN} \,  |F_{(p)}|^{2}\right)  + \\[2mm]
& + & \Big( \frac{1}{2} \, H_{(3)M PQ} H_{(3)N}^{\phantom{(3)M}PQ} - \frac{1}{4} \,  g_{MN} \,  |H_{(3)}|^{2} \Big) \ - \ \Big(4 \nabla_{M} \nabla_{N} \Phi + \frac{1}{2} \, g_{MN} \left(\square \Phi -2\left|\partial \Phi |^2\right.\right)\Big)  \ ,\notag
\end{eqnarray}
where $\nabla_{M}$ is the covariant derivative w.r.t. the Levi-Civita connection. The trace part of the Einstein equation,
\begin{equation}
R\,-\,5|\partial \Phi |^2\,+\,\frac{9}{2} \,\square \Phi \,-\,\frac{1}{4}|H_{(3)}|^2\,-\,\frac{1}{8} \left(e^{2 \Phi }\right) \sum _p (5-p)|F_{(p)}|^2 \, = \, 0 \ ,
\end{equation}

The (modified) BI instead are given by
\be \label{eq:bianchi}
\begin{array}{rclc}
d H_{(3)} & = & 0 & , \\[2mm]
dF_{(0)} & = & 0 & , \\[2mm]
dF_{(2)} \ - \ F_{(0)} \,\wedge\, H_{(3)} & = & 0 & , \\[2mm]
dF_{(4)} \ + \ F_{(2)} \, \wedge \, H_{(3)} & = & 0 & .
\end{array}
\ee
As already noted in the main text, in general all these equations are PDE's. However, in the case of our interest, these reduce to ODE's due to the \emph{Ansatz} \eqref{eq:DWmetric}.
\newpage

\newsection{Half-maximal gauged 7d supergravities} \seclabel{App:Gauged_Sugra}
\noindent In this appendix we introduce half-maximal gauged 7d supergravities. To this end, we will make use of the so-called embedding tensor formalism. 
This tool will allow us to identify a class of deformations which can be shown to correspond with effective 7d theories describing the truncation of massive type IIA supergravity on $S^3$. By focusing on these, we will then review their vacuum solutions. There are only two
distinct and perturbatively stable AdS vacua, a supersymmetric and a non-supersymmetric one, which holographically correspond to a supersymmetric and a non-supersymmetric conformal fixed point, respectively. 

Half-maximal supergravity in $D=7$ coupled to three vector multiplets can be obtained by compactifying type I supergravity on a $\mathbb{T}^{3}$. 
The theory possesses $16$ supercharges which can be rearranged into a pair of symplectic-Majorana (SM) spinors transforming as a doublet of $\textrm{SU}(2)_{R}$. The full Lagrangian enjoys the following 
global symmetry 
\begin{equation}
G_{0} \ = \ \mathbb{R}^{+}_{\phi} \, \times \, \textrm{SO}(3,3) \ \approx \ \mathbb{R}^{+}_{\phi} \, \times \, \textrm{SL}(4,\mathbb{R}) \ .
\end{equation}
The $(64_{B} \ + \ 64_{F})$ bosonic and fermionic propagating degrees of freedom (dof's) of the theory are then rearranged into irrep's of $G_{0}$ as described in table~\ref{Table:dofs}. 
\begin{table}[h!]
\renewcommand{\arraystretch}{1}
\begin{center}
\scalebox{1}[1]{
\begin{tabular}{|c|c|c|c|c|}
\hline
fields & $\textrm{SO}(5)$ irrep's & $\mathbb{R}^{+} \, \times \, \textrm{SL}(4,\mathbb{R})$ irrep's & $\textrm{SU}(2)_{R} \, \times \, \textrm{SU}(2)$ irrep's & \# dof's  \\
\hline \hline
${e_{\mu}}^{a}$ & $\textbf{14}$ & $\textbf{1}_{(0)}$ & $(\textbf{1},\textbf{1})$ & $14$ \\
\hline
${A_{\mu}}^{[mn]}$ & $\textbf{5}$ & $\textbf{6}_{(+1)}$ & $(\textbf{1},\textbf{1})$ & $30$ \\
\hline
$B_{\mu\nu}$ & $\textbf{10}$ & $\textbf{1}_{(+2)}$ & $(\textbf{1},\textbf{1})$ & $10$ \\
\hline
$\phi$ & $\textbf{1}$ & $\textbf{1}_{(+1)}$ & $(\textbf{1},\textbf{1})$ & $1$ \\
\hline
${{\mathcal{V}}_{m}}^{\alpha\hat{\alpha}}$ & $\textbf{1}$ & $\textbf{4}^{\prime}_{(0)}$ & $(\textbf{2},\textbf{2})$ & $9$ \\
\hline
\hline
$\psi_{\mu\alpha}$ & $\textbf{16}$ & $\textbf{1}_{(0)}$ & $(\textbf{2},\textbf{1})$ & $32$ \\
\hline
${\chi}_{\alpha}$ & $\textbf{4}$ & $\textbf{1}_{(0)}$ & $(\textbf{2},\textbf{1})$ & $8$ \\
\hline
$\lambda^{\alpha\hat{\alpha}\hat{\beta}}$ & $\textbf{4}$ & $\textbf{1}_{(0)}$ & $(\textbf{2},\textbf{3})$ & $24$ \\ 
\hline
\end{tabular}
}
\end{center}
\caption{{\it The on-shell field content of (ungauged) half-maximal supergravity in $D=7$. Each field is massless and hence transforms in some irrep of the corresponding little group $\textrm{SO}(5)$ w.r.t.
spacetime diffeomorphisms and local Lorentz transformations. Please note that, in the $\textrm{SL}(4,\mathbb{R})$ scalar coset representative ${{\mathcal{V}}_{m}}^{\alpha\hat{\alpha}}$, one needs to subtract the number
of unphysical scalars corresponding with $\textrm{SO}(4)$ generators in order to come up with the correct number of dof's, \emph{i.e.} $9$.}} \label{Table:dofs}
\end{table}

As one can see from table~\ref{Table:dofs}, the scalar sector of the theory contains an $\mathbb{R}^{+}$ scalar denoted by $\phi$ and an $\frac{\textrm{SL}(4,\mathbb{R})}{\textrm{SO}(4)}$ 
coset representative denoted by $M_{mn}$. In terms of the vielbein ${{\mathcal{V}}_{m}}^{\alpha\hat{\alpha}}$ appearing in table~\ref{Table:dofs}, $M_{mn}$ can be constructed as 
\begin{equation}
M_{mn} \ = \ {{\mathcal{V}}_{m}}^{\alpha\hat{\alpha}} \, {{\mathcal{V}}_{n}}^{\beta\hat{\beta}} \, \epsilon_{\alpha\beta} \, \epsilon_{\hat{\alpha}\hat{\beta}} \ ,
\label{scalarM1}
\end{equation}
where $\epsilon_{\alpha\beta} \, \epsilon_{\hat{\alpha}\hat{\beta}}$ can be viewed as the invariant metric of $\textrm{SU}(2)_{R} \, \times \, \textrm{SU}(2) \ \approx \ \textrm{SO}(4)$.

The kinetic Lagrangian for the scalar sector reads
\begin{equation}
\label{Lkin}
\mathcal{L}_{\textrm{kin}} \ = \ -\frac{1}{2}\,\left(\partial\phi\right)^{2} \ + \ \frac{1}{8}\,\partial_{\mu}M_{mn}\,\partial^{\mu}M^{mn} \ ,
\end{equation}
where $M^{mn}$ denotes the inverse of $M_{mn}$.

It is now possible to deform this theory without breaking supersymmetry by introducing new couplings in the supergravity Lagrangian parametrized by the so-called \emph{embedding tensor} $\Theta$.
Bosonic consistency and supersymmetry restrict $\Theta$ to the following $G_{0}$ irrep's 
\begin{equation}
\begin{array}{lclclclclc}
\Theta & \in & \underbrace{\textbf{1}_{(-4)}}_{\theta} & \oplus & \underbrace{ \textbf{10}^{\prime}_{(+1)}}_{Q_{(mn)}} & \oplus & \underbrace{ \textbf{10}_{(+1)}}_{{\widetilde{Q}}^{(mn)}} & \oplus & 
\underbrace{\textbf{6}_{(+1)}}_{\xi_{[mn]}} & ,
\end{array}
\label{ETirreps}
\end{equation} 
where $\theta$ can be viewed as a St\"uckelberg coupling defining as a so-called $p=3$-type deformation \cite{Bergshoeff:2007vb}, whereas all the other irreducible pieces correspond to traditional gaugings. 
In particular, $Q$ \& $\tilde{Q}$ can be used in order to gauge a subgroup of $\textrm{SL}(4,\mathbb{R})$, whereas $\xi$ necessarily gauges the $\mathbb{R}^{+}_{\phi}$ generator as well as a suitable subgroup of
$\textrm{SL}(4,\mathbb{R})$\footnote{In this 7d theory $\theta$, $Q$ \& $\tilde{Q}$ play the role of F-terms, whereas $\xi$ is the analogue of D-term deformations.}. 

The gauge-invariance of $\Theta$ imposes the following quadratic constraints on the various irreducible components of the embedding tensor
\begin{equation}
\label{QCHalf_Max}
\begin{array}{rclc}
\left(\tilde{Q}^{mp}\,+\,\xi^{mp}\right)\,Q_{pn} \ - \ \frac{1}{4}\,\left(\tilde{Q}^{pq}\,Q_{pq}\right)\,\delta^{m}_{n} & = & 0 & , \\[2mm]
Q_{mp}\,\xi^{pn} \ + \ \,\xi_{mp}\,\tilde{Q}^{pn} & = & 0 & , \\[2mm]
\xi_{mn}\,\xi^{mn} & = & 0 & , \\[2mm]
\theta\,\xi_{mn} & = & 0 & , \\[2mm]
\end{array}
\end{equation}
where $\xi^{mn}\,\equiv\,\frac{1}{2}\,\epsilon^{mnpq}\,\xi_{pq}$. The above constraints contain irreducible pieces transforming in the $\textbf{1}_{(+2)}\,\oplus\,\textbf{6}_{(-3)}\,\oplus\,
\textbf{15}_{(+2)}$ of $\mathbb{R}^{+}_{\phi} \, \times \, \textrm{SL}(4,\mathbb{R})$.

Any solution to the constraints in (\ref{QCHalf_Max}) specifies a consistent deformation of the theory. The classification of $G_{0}$-orbits of consistent deformations was partially carried out in \cite{Dibitetto:2012rk}
and more recently completed in \cite{Dibitetto:2015bia}.

The main consequence of turning on embedding tensor deformations while closing the supersymmetry algebra, is that of inducing a potential term in the Lagrangian for the scalar sector. This reads
\begin{equation}
\label{VHalf_Max}
\hspace{-6mm}
\begin{array}{lcl}
V & = & 2\,\theta^{2}\,e^{-4\sqrt{\frac{2}{5}}\,\phi} \ + \ \frac{1}{2}\,Q_{mn}Q_{pq}\,e^{\sqrt{\frac{2}{5}}\,\phi}\,\left(2M^{mp}M^{nq}\,-\,M^{mn}\,M^{pq}\right) \ +  \\[1mm]
& + & 3\,\xi_{mn}\xi_{pq}\,e^{\sqrt{\frac{2}{5}}\,\phi}\,M^{mp}M^{nq} \ + \ \frac{1}{2}\,\tilde{Q}^{mn}\tilde{Q}^{pq}\,e^{\sqrt{\frac{2}{5}}\,\phi}\,\left(2M_{mp}M_{nq}\,-\,M_{mn}\,M_{pq}\right) \ +  \\[1mm]
& - & 2\,\theta\,\left(Q_{mn}M^{mn}\,-\,\tilde{Q}^{mn}M_{mn}\right)\,e^{-3\phi/\sqrt{10}} \ + \ 2\, Q_{mn}\tilde{Q}^{mn}\,e^{\sqrt{\frac{2}{5}}\,\phi}  \ .
\end{array}
\end{equation}
When restricting to the case of pure F-term-like deformtions (\emph{i.e.} $\xi_{mn}=0$), the potential (\ref{VHalf_Max}) admits a superpotential formulation
\begin{equation}
\label{VfromW}
\begin{array}{lcl}
V & = & \frac{1}{2}\,K^{IJ} \, \partial_{I}W \, \partial_{J}W \, - \, \frac{3}{10} \, W^{2} \ ,
\end{array}
\end{equation}
where $W$ is a real function of the scalars given by
\begin{equation}
\label{Wexpr}
\begin{array}{lcl}
W & \equiv & 2\theta\,e^{-2\sqrt{\frac{2}{5}}\,\phi} \, + \, e^{\phi/\sqrt{10}}\,\left(Q_{mn}M^{mn}\,-\,\tilde{Q}^{mn}\,M_{mn}\right) \ .
\end{array}
\end{equation}

\newsubsubsection{Relation to massive type IIA compactifications} \subseclabel{subsec:ET/fluxes}
In this subsection we will discuss which part of the deformations in (\ref{ETirreps}) can be obtained upon reducing massive type IIA supergravity on $\textrm{AdS}_{7}\times S^{3}$.
In order to derive the embedding tensor/fluxes dictionary for this specific case we will make use of group-theoretical arguments along the lines of \cite{Danielsson:2015tsa}.

This compactifications is generically supported by D6 branes and O6 planes and, as a consequence, it requires a $\mathbb{Z}_{2}$ truncation to the orientifold-even sector, thus resulting in an effective
7d description only retaining $16$ supercharges. As a first check, we will now show how the even components of the IIA bosonic fields fill the 7d degrees of freedom given in table~\ref{Table:dofs}. 
The resulting field content is listed in table~\ref{Table:IIAfields}.
\begin{table}[h!]
\renewcommand{\arraystretch}{1}
\begin{center}
\scalebox{1}[1]{
\begin{tabular}{|c|c|c|}
\hline
IIA fields & $\mathbb{Z}_{2}$-even components & 7d fields  \\
\hline \hline
\multirow{2}{*}{$g_{MN}$} & $g_{\mu\nu}$ & graviton $(\times 1)$\\
\cline{2-3} & $g_{ij}$ & scalars $(\times 6)$ \\
\hline
$B_{MN}$ & $B_{\mu i}$ & vectors $(\times 3)$\\
\hline
$\Phi$ & $\Phi$ & scalar $(\times 1)$\\
\hline
$C_{M}$ & $C_{i}$ & scalars $(\times 3)$\\
\hline
\multirow{2}{*}{$C_{MNP}$} & $C_{\mu\nu\rho}$ & 3-form $(\times 1)$\\
\cline{2-3} & $C_{\mu jk}$ & vectors $(\times 3)$ \\
\hline
\end{tabular}
}
\end{center}
\caption{{\it The effective 7d bosonic field content of type IIA supergravity on a compact 3d manifold. Please note that the orientifold truncation selects the field components which are even w.r.t.
a combination of the fermionic number, the worldsheet parity and O6 involution. After dualizing the 3-form into a 2-form, the counting exactly matches the bosonic part of the 7d field content given in 
table~\ref{Table:dofs}.}} \label{Table:IIAfields}
\end{table}

We introduce the universal moduli $(\rho,\tau)$ \cite{Hertzberg:2007wc} through the following 10d \emph{Ansatz} in the string frame
\be
ds_{10}^{2} \ = \ \tau^{-2} \, g^{(7)}_{\mu\nu} \, dx^{\mu} \otimes dx^{\nu} \, + \, \rho^{2} \, M_{ij} \, dx^{i} \otimes dx^{j} \ ,
\ee
where $M_{ij}$ parametrizes the $\textrm{SL}(3,\mathbb{R})/\textrm{SO}(3)$ coset and the condition
\be
e^{2\Phi} \ \overset{!}{=} \ \rho^{3} \, \tau^{-5} 
\ee
guarantees being in the 7d Einstein frame after reduction. From the 10d viewpoint, all fluxes that can be turned on are naturally irrep's of 
$\mathbb{R}^{+}_{\rho}\times\mathbb{R}^{+}_{\tau}\times\textrm{SL}(3,\mathbb{R})$. This applies to the Romans' mass $F_{(0)}$, the NS-NS gauge flux $H_{(3)}$ wrapping the internal manifold completely, and the
curvature of $S^{3}$ parametrized by the embedding metric $\Theta_{ij}$.

The embedding of $(\rho,\tau)$ within the 7d supergravity scalars reads
\be
\begin{array}{lccclc}
\left\{
\begin{array}{lclc}
\rho & = & e^{\frac{1}{3}\left(2\sqrt{\frac{2}{5}} \, \phi \, - \, \frac{1}{8}\sqrt{\frac{3}{2}} \, \varphi\right)} & , \\[2mm]
\tau & = & e^{\frac{1}{8}\sqrt{\frac{3}{2}} \, \varphi} & ,
\end{array}
\right. & & \textrm{ and } &  & 
M_{mn} \ = \ \left(\begin{array}{c|c}e^{\frac{\varphi}{\sqrt{6}}} \, M_{ij} & \\[2mm] \hline  & e^{-\sqrt{\frac{3}{2}}\varphi}\end{array}\right) & .
\end{array}
\ee
In order to identify the embedding tensor deformations turned on by the type IIA fluxes on the AdS$_{7}\times S^{3}$ background, we need to decompose the objects in (\ref{ETirreps}) w.r.t. the
$\left(\mathbb{R}^{+}\right)^{2}\times\textrm{SL}(3,\mathbb{R})$ subgroup of $G_{0}$. The identification requires the following $\mathbb{R}^{+}$ weight relabeling \cite{Danielsson:2013qfa}
\be
\label{dictionary_weights}
\begin{array}{lclccclclc}
q_{\phi} & = & -\frac{5}{4} q_{\rho} \, - \, \frac{1}{4} q_{\tau} & , &  & q_{\varphi} & = & -\frac{1}{2} q_{\rho} \, + \, \frac{3}{2} q_{\tau} & .
\end{array}
\ee
The explicit embedding of the three phyisical type IIA compact coordinates $x^{i}$ is such that 
\be
\begin{array}{cclc}
\textrm{SL}(4,\mathbb{R}) & \supset & \mathbb{R}^{+}_{\varphi}\times\textrm{SL}(3,\mathbb{R}) \\[4mm]
\textbf{4} & \longrightarrow & \textbf{1}_{(-3)} \, \oplus \, \textbf{3}_{(+1)} & , \\[4mm]
m & \longrightarrow & 4 \, \, \oplus \, \, i & .
\end{array} \nonumber
\ee
This procedure then yields
\be
\begin{array}{cclc}
\mathbb{R}^{+}_{\phi} \, \times \, \textrm{SL}(4,\mathbb{R}) & \supset & \mathbb{R}^{+}_{\phi}\times\mathbb{R}^{+}_{\varphi}\times\textrm{SL}(3,\mathbb{R}) \\[4mm]
\textbf{1}_{(-4)} & \longrightarrow & \textbf{1}_{(-4; \,0)} & , \\[2mm]
\textbf{6}_{(+1)} & \longrightarrow & \textbf{3}_{(+1; \,+2)} \, \oplus \, \textbf{3}^{\prime}_{(+1; \,-2)} & , \\[2mm] 
\textbf{10}_{(+1)} & \longrightarrow & \textbf{1}_{(+1; \,-6)} \, \oplus \, \textbf{3}_{(+1; \,-2)} \, \oplus \, \textbf{6}_{(+1; \,+2)} & , \\[2mm] 
\textbf{10}^{\prime}_{(+1)} & \longrightarrow & \textbf{1}_{(+1; \,+6)} \, \oplus \, \textbf{3}^{\prime}_{(+1; \,+2)} \, \oplus \, \textbf{6}^{\prime}_{(+1; \,-2)} & ,
\end{array} \nonumber
\ee
the resulting embedding tensor/fluxes dictionary being given in table~\ref{table:ET/fluxes}.
\begin{table}[h!]
\begin{center}
\scalebox{1}[1]{
\begin{tabular}{| c | c | c |}
\hline
IIA fluxes & $\Theta$ components  &  $\mathbb{R}^{+}_{\phi}\times\mathbb{R}^{+}_{\varphi}\times\textrm{SL}(3,\mathbb{R})$ irrep's \\[1mm]
\hline \hline
$F_{(0)}$ & $\sqrt{2} \, \tilde{Q}^{44}$ & $\textbf{1}_{(+1; \,-6)}$ \\[1mm]
\hline
$H_{ijk}$ & $\frac{1}{\sqrt{2}} \, \theta \, \epsilon_{ijk}$ & $\textbf{1}_{(-4; \,0)}$ \\[1mm]
\hline
$\Theta_{ij}$ & $Q_{ij}$ & $\textbf{6}^{\prime}_{(+1; \,-2)}$ \\[1mm]
\hline
\end{tabular}
}
\end{center}
\caption{{\it The embedding tensor/fluxes dictionary for the case of massive type IIA reductions on $S^{3}$. The underlying 7d gauging is generically is $\textrm{ISO}(3)$, except when 
$Q_{ij} \, =  \, \mathds{1}_{3}$ and $\tilde{Q}^{44} \, = \, 1$, where it degenerates to $\textrm{SO}(3)$ and supersymmetry gets restored \protect\cite{Louis:2015mka}.} 
\label{table:ET/fluxes}}
\end{table}

By subsequently plugging the mapping in table~\ref{table:ET/fluxes} into the constraints given in \eqref{QCHalf_Max}, one finds that no condition needs to be imposed for consistency. 
This implies that our massive type IIA backgrounds can be effectively described by a theory with sixteen supercharges.
Moreover, one more interesting check consists in working out the extra constraints required in order to have an uplift to maximal supergravity in this case. What one finds is 
$F_{(0)} \, H_{(3)} \, = \, 0$, which is in perfect agreement with the prediction that maximal supersymmetry should not allow for any D6 branes or O6 planes. 

Note that the theory considered in section~\ref{sec:7dGravity} is obtained by the one presented in appendix by performing the following truncation to the gravity sector:
\be
\label{truncation}
\begin{array}{lcclcclc}
M^{mn} \ = \ \delta^{mn} & , & & Q_{mn} \ = \ \textrm{diag}(q,q,q,0) & , & & \tilde{Q}^{mn} \ = \ \textrm{diag}(0,0,0,\tilde{q}) & .
\end{array}
\ee
By specifying \eqref{VHalf_Max} \& \eqref{Wexpr} to the case in \eqref{truncation}, one recovers the scalar potential and the superpotential given in \eqref{V_ISO3} \& \eqref{W_ISO3}, respectively.


\bibliography{references}
\bibliographystyle{utphys}


\end{document}